\documentclass[12pt]{article}
\usepackage{epsfig,here}
\def\tableline{\noalign{
\hrule height.7pt depth0pt\vskip3pt}}

\begin{document}
\setcounter{footnote}{0}
\renewcommand{\thefootnote}{\fnsymbol{footnote}}
\newcommand{\mycomm}[1]{\hfill\break
$\phantom{a}$\kern-3.5em{\tt===$>$ \bf #1}\hfill\break}
\newcommand{\mycommA}[1]{\hfill\break
$\phantom{a}$\kern-3.5em{\tt+++$>$ \bf #1}\hfill\break}
\newcommand{\ksl}{\mbox{$k$\hspace{-0.5em}\raisebox{0.1ex}{$/$}}}
\newcommand{\psl}{\mbox{$p$\hspace{-0.4em}\raisebox{0.1ex}{$/$}}}
\newcommand{\pbsl}{\mbox{$\bar{p}$\hspace{-0.4em}\raisebox{0.1ex}{$/$}}}
\newcommand{\qsl}{\mbox{$q$\hspace{-0.45em}\raisebox{0.1ex}{$/$}}}
\def\beq{\begin{equation}}
\def\eeq{\end{equation}}
\def\MSbar {\hbox{$\overline{\hbox{\tiny MS}}\,$}}
\def\eff{\hbox{\tiny eff}}
\def\res{\hbox{\tiny res}}
\def\FP{\hbox{\tiny FP}}
\def\PV{\hbox{\tiny PV}}
\def\IR{\hbox{\tiny IR}}
\def\UV{\hbox{\tiny UV}}
\def\ECH{\hbox{\tiny ECH}}
\def\NP{\hbox{\tiny NP}}
\def\APT{\hbox{\tiny APT}}
\def\QCD{\hbox{\tiny QCD}}
\def\CMW{\hbox{\tiny CMW}}
\def\spins{\hbox{\tiny spins}}
\def\lsim{\mathrel{\raise.3ex\hbox{$<$\kern-.75em\lower1ex\hbox{$\sim$}}}}
\def\gsim{\mathrel{\raise.3ex\hbox{$>$\kern-.75em\lower1ex\hbox{$\sim$}}}}
\def\MSbar {\hbox{$\overline{\hbox{\tiny MS}}\,$}}
\def\SDG{\hbox{\tiny SDG}}
\def\pinch{\hbox{\tiny pinch}}
\def\brem{\hbox{\tiny brem}}
\def\V{\hbox{\tiny V}}
\def\BLM{\hbox{\tiny BLM}}
\def\NLO{\hbox{\tiny NLO}}
\def\PT{\hbox{\tiny PT}}
\def\PA{\hbox{\tiny PA}}
\def\1loop{\hbox{\tiny 1-loop}}
\def\2loop{\hbox{\tiny 2-loop}}
\def\mysim{\kern -.1667em\lower0.8ex\hbox{$\tilde{\phantom{a}}$}}
\def\a{\bar{a}}

\begin{titlepage}
\begin{flushright}
{\small CERN-TH/2001-371}\\
{\small TSL/ISV-2001-0258}\\
{\small January, 2002}
\end{flushright}

\vspace{.13in}

\begin{center}
{\Large {\bf The thrust and heavy-jet mass distributions \\ in the two-jet region\footnote{
Research supported in part by the EC
program ``Training and Mobility of Researchers'', Network
``QCD and Particle Structure'', contract ERBFMRXCT980194, and the
Swedish Research Council, contract F620-359/2001.
}}}

\vspace{.5in}

{\bf Einan Gardi$^{(1)}$} \,and\, {\bf Johan Rathsman$^{(1,2)}$}

\vspace{0.25in}

$^{(1)}$ TH Division, CERN, CH-1211 Geneva 23, Switzerland\\ 

\vspace{0.15in}

$^{(2)}$ High Energy Physics, Uppsala University,
Box 535, S-751 21 Uppsala, Sweden\\
\vspace{.5in}

\end{center}

\noindent {\bf Abstract:}
Dressed Gluon Exponentiation (DGE) is used to calculate the thrust and the
heavy-jet mass distributions in  $e^+e^-$ annihilation in the two-jet region.
We perform a detailed analysis of power corrections, taking care of the effect
of hadron masses on the measured observables.  In DGE the Sudakov exponent is
calculated in a renormalization-scale invariant way using renormalon
resummation.  Neglecting the correlation between the hemispheres in the two-jet
region, we express the thrust and the heavy-jet mass distributions in terms of
the single-jet mass distribution. This leads to a simple description of the
hadronization corrections to both distributions in terms of a single shape
function, whose general properties are deduced from renormalon ambiguities.  
Matching the resummed result with the available next-to-leading order
calculation, we get a good description of the thrust distribution in a
wide range, whereas the description of the heavy-jet mass distribution, which
is more sensitive to the approximation of the phase space, is restricted to the
range $\rho_H<1/6$.  This significantly limits the possibility to determine
$\alpha_s$ from this observable.  However, fixing $\alpha_s$ by the thrust 
analysis, we show that the power corrections for the two observables are in
good agreement.

\vspace{.25in}

\end{titlepage}

\section{Introduction}

Hadronic jets from the fragmentation of high-energy partons 
is one of the most spectacular phenomena in particle physics. 
It also provides evidence for the partonic nature of hadrons:
observables such as event-shape distributions can be calculated 
perturbatively in terms of quarks and gluons and compared with 
hadronic final state measurements~\cite{Sterman:1977wj}. 
At the same time, these observables are directly sensitive to 
the hadronization process, providing an opportunity to learn 
more about the confinement mechanism from experiment.

The LEP experiments and their high quality data have prompted much 
theoretical progress in the understanding of QCD radiation and in the development of appropriate tools. This includes Sudakov resummation~\cite{Catani:1991kz}--\cite{Dokshitzer:1998kz},
renormalon resummation~\cite{Average_thrust}--\cite{Thrust_distribution}
and parametrization of power corrections~\cite{Average_thrust}--\cite{Salam:2001bd}.

Our goal is to extract from QCD quantitative predictions for
event-shape distributions in the two-jet region. The precise LEP
data provide a strong motivation for such an effort: it can be
used for precision measurements of $\alpha_s$, but most
importantly to test and improve our understanding of the QCD
dynamics. An essential requirement is to
understand how perturbative QCD can provide a reliable starting
point for the calculation of event-shape distributions. Only then can one
address the question of non-perturbative corrections.

The region where perturbation theory applies is limited from both
ends. The multijet region is in principle accessible by going to
ever higher orders. In practice, however,
these calculations soon become difficult. On the other hand, the
cross section is large only in the two-jet region, where it is
determined by soft and collinear radiation. Close enough to the
two-jet limit, the relevant scale for gluon emission becomes of
order of the QCD scale $\Lambda$, and the problem becomes essentially non-perturbative.

In the two-jet region one must take into account the two immediate
consequences of the restricted phase space, namely  
{\em multiple emission} and {\em low scales}. Even a
qualitative description of the distribution ${d\sigma/{dy}\,(Q,y)}$, 
where $Q$ is the centre-of-mass energy and $y$ is an event-shape 
variable that vanishes in the two-jet limit, requires {\em all-order} 
resummation of Sudakov logs $L\equiv\ln 1/y$. Such resummation guarantees 
the vanishing of the cross section at $y=0$, whereas any fixed-order
calculation diverges in this limit.

Sudakov logs can be resummed thanks to the factorization property 
of QCD matrix elements. The minimal
resummation required for quantitative predictions is one where the
Sudakov exponent is calculated to next-to-leading logarithmic
(NLL) accuracy~\hbox{\cite{Catani:1991kz}--\cite{Dokshitzer:1998kz}}.
In NLL resummation the exponent includes terms of the 
form $\alpha_s^nL^{n+1}$ and
$\alpha_s^n L^{n}$ for any~$n$, while terms of the form 
$\alpha_s^nL^{m}$ with $m<n$ are neglected.
The NLL exponentiation formula has become the standard tool 
for data analysis in recent years. Superficially this resummation
applies for arbitrarily small $y$. 
However, due to the running of the coupling it actually applies only for $yQ \gg\Lambda$. In particular, when $yQ\sim \Lambda$ not only does 
the perturbative approximation break down completely, 
but also power corrections of arbitrary power
$\sim \left(\Lambda/(yQ)\right)^n$ become important.

Since the distribution peak is located at $yQ \gsim \Lambda$,
Sudakov resummation with  a fixed logarithmic accuracy, such as
NLL, is insufficient there.  The low scales involved  manifest
themselves in renormalons: large perturbative coefficients that
increase factorially at large orders. These large coefficients
enhance the subleading logs $\alpha_s^nL^{m}$ 
with $m<n$, disrupting the dominance of the leading logs.
Estimates~\cite{Thrust_distribution} of the contribution of
subleading logs, which are neglected  in the standard NLL
resummation, to the thrust ($T$) distribution at $(1-T)Q
\gg\Lambda$ is of the order of $20\%$ at $Q={M_{\rm Z}}$.

In Dressed Gluon Exponentiation
(DGE)~\cite{Thrust_distribution,DGE} the Sudakov exponent is
calculated, starting with an off-shell gluon emission. Following
the spirit of BLM~\cite{BLM}--\cite{BGGR} 
and~\cite{DMW,Average_thrust}, the
actual gluon virtuality is used as the scale of the running
coupling. Through the integral over the coupling DGE resums the
renormalon contribution (in the large $\beta_0$ limit) to the
Sudakov exponent, and thus it takes into account the factorially
enhanced subleading logs to any logarithmic accuracy. As a
consequence, the calculated exponent is renormalization-scale
invariant.

The calculation of the exponent by DGE
makes the simplifying assumption that gluons are emitted independently. 
While reproducing the exact result to NLL accuracy~\cite{Catani:1991kz,Catani:1993ua}, this  simplification amounts to approximations in both the phase space and
the matrix elements beyond this accuracy. Nevertheless, since the two sources of large coefficients, Sudakov logs and renormalons, are accounted for, we
consider DGE a reliable starting point for the analysis of the
distribution in the  range $yQ \gsim \Lambda$. Eventually, 
the dominance of the corrections we resum should be tested when 
the complete next-to-next-to-leading logarithmic (NNLL) and
next-to-next-to-leading order (NNLO) corrections becomes
available.

Physically it is clear that non-perturbative corrections due 
to soft gluon emission are important in the two-jet region. 
Technically, the DGE
calculation gives a clear indication for the presence of
power-suppressed  corrections: the resummation of infrared
renormalons is ambiguous at power accuracy. Consistency of the
full cross section implies the existence of genuine power
corrections  of the same form. 
As in~\cite{Manohar:1995kq,Webber:1994cp} 
and~\cite{Dokshitzer:1995zt}--\cite{Dokshitzer:1997ew} we will assume that 
these power
corrections dominate over power corrections that have no trace in
the perturbative calculation. This has far-reaching implications
concerning the functional form of the non-perturbative correction.
The leading corrections can be written as a convolution of the
perturbative result with a shape function of a {\em single}
variable $yQ$, as suggested by Korchemsky and 
Sterman~\cite{Korchemsky:1995is}--\cite{Belitsky:2001ij},~\cite{Thrust_distribution}.
The simplest approximation to the shape function amounts to a
shift of the perturbative distribution, in accordance with~\cite{Shape_function2,Dokshitzer:1997ew}. 
Based on the renormalon ambiguity (in
the large $\beta_0$ limit) only power corrections corresponding to
odd central moments of the shape function are present. This gives,
on the one hand, an explanation for the success of shift-based fits
and, on the other hand, an opportunity to test the renormalon
dominance assumption by fitting the data and extracting higher
moments.

The DGE method has been the basis of our analysis of the thrust
distribution in~\cite{Thrust_distribution}. The comparison with
experimental data was encouraging in several respects: first, good
fits were obtained in a wide range of energies and thrust values.
These fits were quite stable under a variation of different
parameters, such as the cuts on the fitting range in thrust. In
addition, the value of $\alpha_s$ and the  non-perturbative
parameter were consistent between the shape-function- and the 
shift-based fits. Moreover,  these values were also consistent with those
obtained by fitting the average thrust using  renormalon
resummation~\cite{Average_thrust}.  Such consistency, which is a
crucial test of whether the approximation used is satisfactory,
is not achieved~\cite{MovillaFernandez:2001ed} in standard NLL fits 
where renormalons are not resummed.

One aspect that certainly deserves attention is the fact that the
best fit value of $\alpha_s$ obtained
in~\cite{Thrust_distribution,Average_thrust},
$\alpha_s^{\MSbar}({M_{\rm Z}})=0.110$, is significantly lower than
the world average~\cite{PDG} value $\alpha_s^{\MSbar}({M_{\rm Z}})=0.1172\pm
0.0020$. The experimental uncertainty is very small and the
sources of theoretical uncertainty investigated in~\cite{Thrust_distribution}
hardly sum up to~$5\%$. The latter uncertainty is dominated by
non-logarithmic corrections at NNLO, which should become available
in the foreseeable future.

It is of great interest to extend the application of DGE to more
event-shape distributions. The main motivation is to
test and improve our understanding of various perturbative and
non-perturbative questions. In particular, it is 
interesting to investigate the relations between non-perturbative
corrections to different observables. The idea of a
``universal'' (observable-independent) origin of the leading power
corrections to various event-shape observables has been the subject
of much theoretical and phenomenological activity in the past few
years. Universal power corrections were suggested both within the infrared
finite coupling
approach~\cite{Dokshitzer:1995zt}--\cite{Dokshitzer:1998pt},
and in the shape-function approach~\cite{Korchemsky:2000kp}. In
either case, the observable independence of the power corrections
involves additional assumptions.

Probably the most relevant pair of variables to be compared in this respect 
are the thrust and the heavy-jet mass. For completeness we  recall their
definitions here, together with some other, closely related, jet mass variables:
\begin{eqnarray}
\mbox{Thrust} &&
T\equiv\max_{\vec{n}_T} \frac{\sum_i|\vec{p}_i \cdot \vec{n}_T|}{\sum_i|\vec{p}_i|}
\label{eq:tdef} \\
\mbox{Single-jet mass} &&
\rho_m\equiv\frac{ \left(\sum_{i\in {\rm H}_m} p_i \right)^2}{\left(\sum_i p_i \right)^2 }
=\frac{ \left(\sum_{i\in {\rm H}_m} p_i \right)^2}{Q^2 }
\label{eq:rdef} \\
\mbox{Heavy-jet mass} &&
\rho_H\equiv\max(\rho_1,\rho_2)
\label{eq:hdef} \\
\mbox{Light-jet mass} &&
\rho_L\equiv\min(\rho_1,\rho_2).
\label{eq:ldef}
\end{eqnarray}
Here the index $i$ runs over all the particles and $m=1, 2$ represents
the two hemispheres ${\rm H}_m$ defined by the thrust axis $\vec{n}_T$.

There are several useful relations between these observables,
which follows from their definition:
the single-jet mass is the average of the heavy and the light,
\begin{equation}
\frac{1}{\sigma}\frac{d\sigma}{d\rho} = \frac{1}{2}
\left[ \frac{1}{\sigma}\frac{d\sigma}{d\rho_H}
+\frac{1}{\sigma}\frac{d\sigma}{d\rho_L} 
\right] \; .
\label{eq:jetrel}
\end{equation}
In the two-jet region $t\equiv 1-T$ equals, up to corrections 
suppressed by $t$, the sum of the two jet masses~\cite{Catani:1993ua}
\begin{equation}
t\simeq \rho_H+\rho_L \; .
\end{equation}
To leading order in $\alpha_s$ this relation is exact and, since
the light-jet mass vanishes at this order, it follows
that
\begin{equation}
\frac{1}{\sigma}\frac{d\sigma}{dt} =
\frac{1}{\sigma}\frac{d\sigma}{d\rho_H} = 2
\frac{1}{\sigma}\frac{d\sigma}{d\rho} \quad  \quad
\left({\cal O}(\alpha_s) \right).
\label{eq:lorel}
\end{equation}
In the approximation of independent emission used here, both the
thrust and heavy-jet mass
distributions can be expressed in terms of the single-jet mass
distribution~\cite{Catani:1993ua}.  This implies that a single 
shape function~\cite{Korchemsky:2000kp} can be used to parametrize the
non-perturbative corrections in both cases. On the other hand,
fits based on the NLL resummation formula and a shift for the two
distributions yield different values of $\alpha_s$, along with a
different magnitude for the power correction (the
shift)~\cite{MovillaFernandez:2001ed}.  While a fair comparison of
the power corrections must assume a common value of $\alpha_s$,
such a fit does not  improve the agreement between the power
correction parameters.  We will see below (Fig.~\ref{fig:correlation}) that 
fitting the heavy-jet mass distribution by shifting the NLL distribution, as
in~\cite{MovillaFernandez:2001ed}, but {\em fixing} the value of~$\alpha_s$ 
based on the thrust fit, the  power correction of the heavy-jet mass is {\em
much smaller}\footnote{The NLL fits  in~\cite{MovillaFernandez:2001ed} yield a
lower value of $\alpha_s$ for the heavy-jet mass  and a higher value of the
power correction (shift) compared to the thrust values.} ($\sim 50\%$) than
expected based  on the thrust fit.

Since the relation between the power corrections to the heavy-jet mass and those to the thrust follows\footnote{One does not rely here on the assumption of
universality of an infrared finite coupling.} from the definition of these
variables, the discrepancy between the power corrections must be interpreted 
as a sign that the approximations made in the calculation are too crude.  Based
on~\cite{Thrust_distribution,Average_thrust}, it is quite clear that one reason
for this discrepancy is the fact  that renormalon contributions were neglected.
We will verify this assertion in what follows.

In~\cite{Korchemsky:2000kp}, the relation between the
non-perturbative corrections to the two distributions was
explained in terms of a strong correlation between the two
hemispheres due to non-perturbative soft gluon emission at large
angles. It was shown that introducing a positive correlation through a
two-hemisphere shape function, the heavy-jet mass and the  thrust
(as well as the $C$-parameter) can be fitted. However, such a
correlation, if important, should appear first at the
perturbative level. Since the resummation formulae implicitly
assume independent emission, and thus no correlation, we would rather not introduce such a correlation on the
non-perturbative level. We shall assume here that in the two-jet region 
the jet masses are uncorrelated also on the non-perturbative level, and confront this assumption with the data.

Another important issue in the comparison of power corrections to
the thrust and the heavy-jet mass is the dependence of the
distribution on the hadron masses~\cite{Salam:2001bd}. The
perturbative calculation is based on {\em massless} partons, whereas
the measured observable is based on {\em massive} hadrons.  It is
clear that the spectrum of hadrons has some effect on the
distribution. In general at high enough $Q$ and
$y$, the details of the spectrum do not matter. Instead, some
generic characteristics of hadronization dominate, and a small set
of non-perturbative parameters controlling power corrections is
sufficient. However, the comparison of the thrust
and the heavy-jet mass is problematic since the thrust is
(conventionally) defined in terms of the three-momenta whereas the
heavy-jet mass in terms of the four-momenta. This implies that the
two observables will exhibit {\em different sensitivity} to the
masses of the measured particles. Ways to overcome this difficulty
by appropriate modifications of the definitions (which coincide
with the standard ones in the massless limit) were suggested
in~\cite{Salam:2001bd}. We will see that choosing a common modification
of the definition (hadron mass scheme) for the two observables is indeed 
crucial for a meaningful comparison of the power corrections.

The outline of this paper is as follows. We start in section~2  by
recalling the perturbative calculation of the thrust and the heavy
jet mass distribution using DGE. We then discuss the limitations of the approximation at hand, and, finally, we
present our assumptions concerning hadronization corrections.
Section~3 summarizes the data analysis. We begin by performing
separate fits to the thrust  and the heavy-jet mass distributions
and then study the consistency between the two. Section~4 contains
our conclusions.


\section{Assumptions and calculation}

In this section we present the calculation of the distributions, emphasizing
the main assumptions involved. In section~2.1 we calculate the jet mass
distribution from a single dressed gluon (SDG). In section~2.2, we exponentiate
the SDG cross section arriving at the DGE formula for the single-jet mass
distribution. This distribution is then used to express the thrust and the
heavy-jet mass distributions. In section~2.3 we discuss the limitations of the
considered approximation. Finally, in section~2.4 we present our approach
concerning the parametrization of hadronization corrections and the way we 
deal with hadron mass effects.

\subsection{Single Dressed Gluon cross section}

We begin by calculating the differential cross section for a
single gluon emission. The calculation is performed with an
off-shell gluon $k^2=m^2\geq 0$ in order to provide the basis for
renormalon resummation~\cite{DMW,Beneke:1995qe,Ball:1995ni,Average_thrust,Gardi:2000yh,Thrust_distribution}.
Contrary to~\cite{Gardi:2000yh,Thrust_distribution}, which used the 
full matrix element and identified the log-enhanced terms prior to 
the integration over the gluon virtuality, we will follow the approximation 
suggested in~\cite{DGE}, which allows to isolate the relevant terms right 
at the beginning.

By performing the calculation with an off-shell gluon we will
effectively be integrating  inclusively over the decay products of
the gluon. Based on the definitions of the variables
(\ref{eq:tdef}) -- (\ref{eq:ldef}) such a calculation yields the
exact result when the partons that the gluon decays into are
confined to a single hemisphere, but not so when these partons end
up in different hemispheres. Therefore, our result should be
considered as an approximation {\em even} in the large $\beta_0$ (large
$N_f$) limit~\cite{Nason:1995hd}. There are strong indications
that this inclusive approximation is good in the case of the
thrust~\cite{DMW,Average_thrust,Gardi:2000yh,Thrust_distribution}.
We will return to discuss this point in the section~2.3.

Let us denote the final momenta of the two
quarks by $p$ and $\bar{p}$, where $p^2=\bar{p}^2=0$.
Sudakov logs are associated with the singularity of the quark
propagator prior to the emission, $(\psl+\ksl)/(p+k)^2$.
Concentrating in the two-jet region, we will neglect
non-logarithmic terms in the cross section. This is equivalent to
neglecting non-singular terms in the matrix element~\cite{DGE}. It
is convenient to use the light-cone  axial gauge~\cite{DGE} where,
in the approximation considered, the gluon effectively decouples
from one of the quarks.

Assuming that the gluon is in the hemisphere of the quark $p$, it
is $\bar{p}$ that sets the thrust axis (we will calculate the cross section
only in this half of phase space and account for the other half by an 
overall factor of 2). We define the light-cone
coordinates such that $p=(p_{+},0,0)$ and
$\bar{p}=(0,\bar{p}_{-},0)$, and use the light-cone gauge where
$A_+=0$. In this gauge the gluon coupling to the $\bar{p}$ quark
is suppressed\footnote{In the on-shell case, $m^2=0$, this
coupling is strictly zero: $A_+J_{-}=0$.}, so it is enough to
consider the diagram where the gluon is emitted from $p$. The
gluon propagator is given by
\begin{eqnarray}
\label{propagator}
d_{\mu \nu}=-g_{\mu \nu}+\frac{k_{\mu}\bar{p}_\nu+\bar{p}_{\mu}k_\nu}{k\bar{p}}.
\end{eqnarray}

The cross section in the centre-of-mass frame is
\begin{eqnarray*}
\sigma=\frac{e^4 N_c \sum_q e_q^2 }{2Q^2}\int
\frac{d^4p\,d^4\bar{{p}}}{(2\pi)^6}\,\delta\left(p^2\right)\,\delta\left(\bar{p}^2\right)
\frac{d^4k}{(2\pi)^3}\,\delta\left(k^2-m^2\right)(2\pi)^4\delta^4(q-p-\bar{p}-k)
\, \frac{H_{\rho\sigma} \, L^{\rho\sigma}}{Q^4}.
\end{eqnarray*}
The leptonic tensor $L^{\rho\sigma}=\frac14 {\rm
Tr}\left\{\qsl_1\gamma^{\rho} \qsl_2 \gamma^{\sigma}\right\}$ can
be replaced by $\frac13 Q^2 g^{\rho\sigma}$. To calculate the
hadronic tensor $g_{\rho\sigma}\,H^{\rho\sigma}=\sum_{\mbox {\rm
{\spins}}} {\cal{M}}{\cal{M}}^\dagger$ we write the amplitude
\begin{equation}
\label{Amp}
{\cal M}=g_s\,t^a\,{\epsilon^{\lambda}_{\mu}}^*\,\frac{1}{(k+p)^2}\,
\bar{u}^{(s)}(p)\gamma^\mu(\psl+\ksl)\gamma^{\rho}u^{(\bar{s})}({\bar{p}})
\end{equation}
where $t^a$ is a colour matrix and ${\epsilon^{\lambda}_{\mu}}^*$ is
the gluon polarization vector. The squared matrix element summed
over the gluon and quark spins is then given by
\begin{eqnarray}
\label{M2} \sum_{\mbox {\rm {\spins}}} {\cal{M}}{\cal{M}}^\dagger
= \frac{C_F g_s^2}{(k+p)^4}\, g_{\rho\sigma}\,d_{\mu\nu} {\rm
Tr}\left\{(\psl+\ksl)\gamma^\nu \psl \gamma^\mu
(\psl+\ksl)\gamma^{\rho}\pbsl\gamma^{\sigma} \right\}
\end{eqnarray}
where $\sum_{s} u^{(s)}(p)\bar{u}^{(s)}(p)$ was replaced by $\psl$, and
$\sum_{\lambda}{\epsilon_\nu^{\lambda}}^*\epsilon_\mu^{\lambda}$  by
$d_{\mu\nu}$. Introducing the Sudakov decomposition of
$k=(k_{+},k_{-},k_\perp)=\beta p+\alpha \bar{p}+k_\perp$, and defining the following
dimensionless parameters
\begin{eqnarray}
\label{lc_par}
\beta&=&k_{+}/p_{+}=2k\bar{p}/2p\bar{p}\nonumber \\
\alpha&=&k_{-}/\bar{p}_{-}=2kp/2p\bar{p}\\ \nonumber
\gamma&\equiv& k_\perp^2/2p\bar{p} \\ \nonumber
\lambda&=&m^2/2p\bar{p}=(2k_{+}k_{-}-k_\perp^2)/2p\bar{p}=\alpha\beta-\gamma,
\end{eqnarray}
the propagator is proportional to $1/(\alpha+\lambda)^2$. As
observed in~\cite{DGE}, to calculate the log-enhanced
cross section, it is sufficient to take into account the singular
terms in the matrix element, corresponding to the situation where
{\em both} $\alpha$ and $\lambda$ are small (with no specific
hierarchy between them). From the condition
$\alpha\beta=\gamma+\lambda$ it follows that the transverse
momentum fraction $\gamma$ is also small (at least with respect to
$\beta$) while $\beta$ can be either large (the collinear limit)
or small (the soft limit). In these variables
\begin{eqnarray}
\label{M22} \sum_{\mbox {\rm {\spins}}} {\cal{M}}{\cal{M}}^\dagger
= 8C_F g_s^2 \,\left[
\frac{-\lambda}{({\alpha+\lambda})^2}\,(1+\beta)+\frac{1}{\alpha+\lambda}\,\frac{\beta^2+2\beta+2}{\beta}
\right].
\end{eqnarray}
The cross section is therefore\footnote{An overall factor of $2$ accounting
for the other half of phase space is included.},
\begin{eqnarray}
\label{cross-section} \sigma/\sigma_0=\frac{C_F\alpha_s}{\pi}
\int \frac{ d\alpha
\,d\beta\,d\lambda\,\delta(\lambda-\epsilon(1+\beta))}{(1+\beta)^3}
\,\left[
\frac{-\lambda}{({\alpha+\lambda})^2}\,(1+\beta)+\frac{1}{\alpha+\lambda}\,\frac{\beta^2+2\beta+2}{\beta}
\right]
\end{eqnarray}
where $\sigma_0= { N_c (\sum_q e_q^2)\, 4\pi\alpha^2}/{(3Q^2)}$
and the $\delta$ function guarantees that the integration is
performed for a fixed $\epsilon\equiv m^2/Q^2$ and $Q^2$.
The integrand has the interpretation of a generalized, off-shell gluon 
splitting function~\cite{DGE}.

We are now ready to calculate the differential cross section for the jet mass
distribution. The hemisphere of the quark $p$ to which the gluon is emitted
is the heavy one, \mbox{$m_H^2=(p+k)^2/Q^2$}, whereas the other has zero mass,
$m_L^2=0$. The relation between $\rho_H={m_H^2}/{Q^2}$ and the
Sudakov variables~(\ref{lc_par}) is simply 
\beq
\rho_H=\frac{(p+k)^2}{Q^2}=\frac{2pk+k^2}{Q^2}=\frac{(\alpha+\lambda)\,2p\bar{p}}{Q^2}
=\frac{\alpha+\lambda}{1+\alpha+\beta+\lambda}\simeq\frac{\alpha+\lambda}{1+\beta}.
\label{rho_x} 
\eeq
Note that also $t=1-T=(m_H^2+m_L^2)/Q^2=\rho_H$.

Changing variables in (\ref{cross-section}) we get the following
result for the single off-shell gluon differential cross section
\beq 
\frac{1}{\sigma}\frac{d\sigma}{d\rho_H}\,=\,\frac{C_F
\alpha_s}{\pi}\,\int \,\frac{d\beta}{(1+\beta)^2}\,\left[-
\frac{\epsilon}{\rho_H^2}+\frac{1}{\rho_H}\,\frac{\beta^2+2\beta+2}{\beta\,(1+\beta)}
\right]. \label{splitting_function} 
\eeq

Next, in the integration over $\beta$, it is the limit of vanishing
transverse momentum $\alpha\beta=\lambda$ that generates the
log-enhanced terms~\cite{DGE}. We therefore integrate over $\beta$
and substitute the limit $\beta=\epsilon/(\rho_H-\epsilon)$. The
result is 
\beq 
\frac{1}{\sigma}\frac{d\sigma}{d\rho_H}\,=
\,\frac{C_F \alpha_s}{\pi}\,\left[ \frac2\rho_H \,\ln \frac{\rho_H}{\epsilon}-
\frac32\frac1\rho_H+\frac\epsilon{\rho_H^2}+\frac12\frac{\epsilon^2}{\rho_H^3}
\right]. 
\label{SDG_dcs} 
\eeq

Using the gluon bremsstrahlung effective charge~\cite{Catani:1991rr} $\alpha_s
\longrightarrow \bar{\alpha}_s$ is sufficient to guarantee that the {\em
singular terms} in the splitting functions are {\em exact}. This way the
renormalization scheme is not arbitrary: there is a unique effective charge
that generates correctly subleading terms. Provided one also uses the
two-loop renormalization group equation, the calculation of the thrust and the
heavy-jet mass distributions will be
exact~\cite{Catani:1991rr,Thrust_distribution} to NLL accuracy.

We now replace $\bar{\alpha}_s$ by an integral over the time-like
discontinuity of the coupling~\cite{DMW,Beneke:1995qe,Ball:1995ni}.
In this way we take into account the relevant
physical scale for the gluon emission $k^2 = \epsilon Q^2$. 
Defining $\bar{A}(k^2)\equiv \beta_0\bar{\alpha}_s(k^2)/\pi$,
we will use here the scheme invariant Borel
representation~\cite{Brown:1992pk}--\cite{Grunberg:1993hf}, \beq \bar{A}(k^2)=\int_0^{\infty}d{u}
\,\exp\left({-{u} \ln k^2/\bar{\Lambda}^2}\right) \bar{A}_B({u}),
\label{A_Borel} \eeq
where $\bar{A}_B(u)=1$ for the one-loop coupling, provided $\bar{\Lambda}$ is
defined appropriately. The integrated time-like discontinuity is~\cite{Brown:1992ic,DMW}
\beq
\bar{A}_{\eff} (
Q^2)=\int_0^{\infty}d{u}\,\exp\left(-{{u} \ln
Q^2/\bar{\Lambda}^2}\right)\,
 \frac{\sin\pi{u}}{\pi{u}}\, \bar{A}_B({u}).
\label{A_eff_Borel}
\eeq
After integration by parts
the single {\em dressed} gluon cross section~is \beq
\left.\frac{1}{\sigma}\frac{d\sigma}{d\rho_H}(Q^2,\rho_H)\right\vert_{\SDG}=
\frac{C_F}{\beta_0}\int_{\rho_H^2}^{\rho_H}\frac{d\epsilon}{\epsilon}\,\bar{A}_{\eff}
(\epsilon Q^2)
\left[\frac{2}{\rho_H}-\frac{\epsilon}{\rho_H^2}-\frac{\epsilon^2}{\rho_H^3}\right].
\label{ours_H} \eeq
Here, the upper integration limit is determined
from the condition that $\alpha$ is positive. Along the zero
transverse momentum boundary of phase space, the limit is
$\alpha=(\rho_H-\epsilon)(1+\beta)=0$. In this situation both
$k_\perp^2$ and $k_-$ vanish, so the gluon is exactly collinear. The
lower integration limit is determined  from the requirement that
$\beta>\alpha$ (otherwise the gluon belongs to the $\bar{p}$
hemisphere). Along the zero transverse momentum
boundary~$\alpha\beta=\lambda$, this condition translates into
$\beta>\sqrt{\lambda}$. It follows that the lower integration limit 
is $\epsilon=\rho_H^2$. This limit corresponds
to large-angle, soft emission.

To obtain the single-jet mass distribution we have to average over
the heavy and the light hemispheres. But since the light-jet mass
cross section starts at two gluon emission (order $\alpha_s^2$)
this amounts, according to (\ref{eq:lorel}), to dividing the heavy
jet mass cross section by~2, 
\beq
\left.\frac{1}{\sigma}\frac{d\sigma}{d\rho }(Q^2,\rho
)\right\vert_{\SDG}= \frac{C_F}{2\beta_0}\int_{\rho ^2}^{\rho
}\frac{d\epsilon}{\epsilon}\,\bar{A}_{\eff} (\epsilon Q^2)
\left[\frac{2}{\rho }-\frac{\epsilon}{\rho
^2}-\frac{\epsilon^2}{\rho ^3}\right]. 
\label{ours} 
\eeq
Substituting (\ref{A_eff_Borel}) into (\ref{ours}) and performing 
the integral over~$\epsilon$ yields the Borel
representation of the logarithmically enhanced SDG cross section,
\beq
\left.\frac{1}{\sigma}\frac{d\sigma}{d\rho}(Q^2,\rho)\right\vert_{\SDG}=
\frac{C_F}{2\beta_0}\int_0^{\infty}\,du\, B_{\SDG}(u,\rho)
\,\exp\left(-{{u} \ln
Q^2/\bar{\Lambda}^2}\right)\,\frac{\sin \pi u }{\pi u} \bar{A}_B(u)  
\label{Borel_rep}
\eeq
with the following Borel function~\cite{Thrust_distribution}
\beq
B_{\SDG}(u,\rho)= \frac1\rho\left[\frac{2}{u} \exp\left(2u\ln\frac 1\rho\right)
-\left(\frac{2}{u}+\frac{1}{1-u}+\frac{1}{2-u}\right)
 \exp\left({u} \ln\frac 1\rho\right)\right] \, .
\label{Bf}
\eeq
This result was discussed in detail in~\cite{Thrust_distribution}. Here we
just recall that the sensitivity to soft gluon emission does not appear here
through Borel singularities (which cancel out by the $\sin(\pi u)$ factor) but
rather as convergence constraints on the Borel integration at 
$u\to\infty$.

\subsection{Exponentiation}

Assuming independent emission and additive contribution by each
emission to the jet mass: $\rho=\sum_{k=1}^{n}\rho_k$, the
multiple gluon phase space factorizes in Laplace
space~\cite{Catani:1993ua}. As in~\cite{Thrust_distribution}, 
the DGE cross section is obtained by 
exponentiating the SDG cross section, given by~(\ref{Borel_rep}) 
and (\ref{Bf}), under the Laplace transform. The Laplace transform 
$J(\nu,Q^2)$ of the single-jet mass distribution is then given by 
\beq 
\ln J(\nu,Q^2)=\int_0^1 \left.
\frac{1}{\sigma}\frac{d\sigma}{d\rho}(Q^2,\rho)\right\vert_{\SDG}
\left(e^{-\nu \rho}-1\right)d\rho, \label{ln_J_nu} 
\eeq 
where the Laplace weight $e^{-\nu \rho}$ is associated with real emissions 
and $-1$ with virtual corrections. The corresponding Borel
representation~\cite{Thrust_distribution} is \beq \ln
J(\nu,Q^2)= \frac{C_F}{2\beta_0}\int_0^{\infty}\,du\, B(\nu,u)
\,\exp\left(-{{u} \ln Q^2/\bar{\Lambda}^2}\right)\,\frac{\sin \pi
u }{\pi u} \bar{A}_B(u) \label{J_nu_Borel} \eeq with
\begin{eqnarray}
\label{Borel_nu}
B(\nu,u)&=& \int_0^1 \frac{d\rho}{\rho} \left[\frac{2}{u}
e^{2u \ln\frac1\rho}-\left(\frac{2}{u}+\frac{1}{1-u}+\frac{1}{2-u}\right)
 e^{u \ln\frac1\rho} \right]\,\left(e^{-\nu \rho}-1\right)\\
&\simeq&\frac{2}{u}\left[e^{2u\ln\nu}\Gamma(-2u)
+\frac{1}{2u}\right]-\left(\frac{2}{u}+\frac{1}{1-u}
+\frac{1}{2-u}\right)\left[e^{u\ln \nu}\Gamma(-u)
+\frac{1}{u}\right],\nonumber
\end{eqnarray}
where non-logarithmic terms have been neglected.  Contrary to the SDG result
(\ref{Bf}), the DGE result has a rich renormalon structure, notably, it has
singularities at half-integer values of $u$. These Borel singularities
indicate power corrections. Note that the first term in~(\ref{Borel_nu}),
which has a double argument $2u$, corresponds to large-angle soft emission,
whereas the second corresponds to collinear emission.

The single-jet mass distribution  can be obtained by the inverse
Laplace transform of~(\ref{J_nu_Borel}):
\begin{eqnarray}
\label{eq:sjm}
\frac{1}{\sigma}\frac{d\sigma}{d\rho}(\rho,Q^2)
=\frac{d}{d\rho}\int_C\frac{d\nu}{2\pi i
\nu}\,\exp\left\{\nu \rho+    \ln J(\nu,Q^2)\right\}
\equiv \frac{d}{d\rho} R_{\rho}(Q^2,\rho)\equiv j(\rho,Q^2),
\end{eqnarray}
where the contour $C$ goes from $-i\infty$ to $+i\infty$ to the
right of all the singularities of the integrand. Using this
distribution, the thrust $t=\rho_1+\rho_2$ distribution is
\begin{eqnarray}
\label{eq:t}
\frac{1}{\sigma}\frac{d\sigma}{dt}(t,Q^2)&=&
\int d\rho_1 d\rho_2 \, j(\rho_1,Q^2) j(\rho_2,Q^2)\,
\delta(\rho_1+\rho_2-t)\\
&=&\frac{d}{dt}\int_C\frac{d\nu}{2\pi i
\nu}\,\exp\left\{\nu t+  2  \ln J(\nu,Q^2)\right\}
\equiv \frac{d}{dt} R_{t}(Q^2,t),
\nonumber
\end{eqnarray}
whereas the heavy-jet mass $\rho_H=\rho_1\Theta(\rho_1-\rho_2)+\rho_2\Theta(\rho_2-\rho_1)$
distribution is
\begin{eqnarray}
\label{eq:hjm}
&&\!\!\!\!\!\!\!\!\!\!\!\!\!\!\!\!\!\!\!\!\!\!\!\!
\frac{1}{\sigma}\frac{d\sigma}{d\rho_H}(\rho_H,Q^2)= \nonumber \\
&=&\int
d\rho_1  d\rho_2  \,j(\rho_1,Q^2)
 j(\rho_2,Q^2)\,\left[\delta(\rho_1-\rho_H)\Theta(\rho_1-\rho_2)+
\delta(\rho_2-\rho_H)\Theta(\rho_2-\rho_1)
\right]\nonumber \\
&=&\frac{d}{d\rho_H}\,\int
d\rho_1  d\rho_2  \,j(\rho_1,Q^2)
 j(\rho_2,Q^2)\,\Theta(\rho_H-\rho_1)\Theta(\rho_H-\rho_2)\nonumber \\
&=&\!\!\!\!\frac{d}{d\rho_H}\,
\left[\int_C\frac{d\nu}{2\pi i\nu}
\,\exp\left\{\nu\rho_H+ \ln J(\nu,Q^2)\right\}\right]^{ 2 }
\equiv \frac{d}{d\rho_H}\, R_{\rho_H}(Q^2,\rho_H).
\end{eqnarray}

\setcounter{footnote}{0} Equations (\ref{eq:t}) and~(\ref{eq:hjm})
are our final expressions (prior to matching with the NLO result)
for the thrust and the heavy-jet mass distributions. Using $\ln
J(\nu,Q^2)$ in (\ref{J_nu_Borel}) with the two-loop gluon
bremsstrahlung coupling, as we do below, these results are exact
to NLL accuracy. However, they differ from~\cite{Catani:1993ua} by
large subleading logs which emerge from the integral over the
running coupling. In order to understand how significant these
additional terms are, it is useful~\cite{Thrust_distribution} to
rewrite $\ln J(\nu,Q^2)$ as an asymptotic series in the
coupling,
\begin{eqnarray}
\ln J(\nu,Q^2)  & =  & \frac{C_F}{2\beta_0}
 \sum_{k=1}^{\infty} f_k\left( \xi \right)
  \bar{A}(Q^2) ^{k-2} \; ,
\label{eq:spt_fixed}
\end{eqnarray}
where $\xi=\bar{A}(Q^2)\ln\nu$. Truncation of this series amounts to a
fixed logarithmic accuracy. The first two functions ($k=1 ,2$)
coincide with the known leading and next-to-leading log functions, whereas
$f_k$ for $k\geq 2$ are given by
\begin{eqnarray*}
\begin{array}{ll}
f_{3}(\xi)=   { 0.804}/{(1-\xi ) }  - {  {2.32}/({1 -2\xi   })}  \\
f_{4}(\xi)= {0.779}/{({1-\xi  } )^{2}} - {  {2.68}/{(1 -2\xi )^{2}}}  \\
f_{5}(\xi)=   { 2.32}/{({1-\xi  })^{3}}  - {  {5.00}/{(1 -2\xi )^{3}}}
\end{array}
\end{eqnarray*}
where, for simplicity, we used a one-loop running coupling.
At higher orders, these functions behave as
\begin{eqnarray}
f_k(\xi)  & \sim   & (k-3)!\,\left[ \frac{1}{(1-\xi)^{k-2}} -
2.5\,\frac{1}{(1-2\xi)^{k-2}}\right].
\end{eqnarray}
The factorially increasing coefficients of $f_k(\xi)$ and the
enhanced singularity at $\xi=1/2$ imply that a fixed logarithmic
accuracy calculation holds only in a very restricted range in the
$\bar{A}(Q^2)$  and $L=\ln\nu\simeq \ln 1/\rho$ parameter space.
Note that the contributions of two consequent terms in
(\ref{eq:spt_fixed}) are of the same order when
$\bar{A}(Q^2)\simeq (1-2\xi)/(k-2)$. At this order the
perturbative expansion is exhausted (the minimal term is reached)
and the correction is effectively power-like  $\sim
\,\bar{\Lambda}/(Q\rho)$.  We assume below that the order of the
minimal term is higher than NLL, so including higher orders in
(\ref{eq:spt_fixed}) improves the evaluation of  $\ln
J(\nu,Q^2)$. Figures~2 and~3 in \cite{Thrust_distribution} show
that near the distribution peak, and anywhere above it,
$\rho>\rho({\rm peak})$, this is indeed the case.
The fixed logarithmic accuracy calculation is justified 
if~(\ref{eq:spt_fixed}) converges well. 
For a fixed $\xi$ (any $\xi<1/2$) this holds if the coupling is very small $\bar{A}(Q^2)\ll 1-2\xi$. However, for a given coupling, there is good convergence only for $\xi\ll 1/2$,~i.e.
\[
{\rho} \gg \exp\left(-\frac{1}{2\bar{A}(Q^2)}\right)=\frac{\bar{\Lambda}}{Q}.
\]
This condition is not realized in the peak region at any relevant
centre-of-mass energy. Consequently a fixed logarithmic accuracy
is insufficient. The analysis
of~\cite{Thrust_distribution} showed, that in the case of the
thrust, the contributions of NNLL and further subleading logs to
$\ln J(\nu,Q^2)$ at $Q={M_{\rm Z}}$ lead to a correction of $\sim
20\%$ for $\rho>\rho({\rm peak})$. Only when all the terms (up to
the minimal term) in the series (\ref{eq:spt_fixed}) are kept can
one safely combine the perturbative calculation with the
corresponding power corrections, eventually extending the
applicability of the result to the peak region.

As explained in~\cite{Thrust_distribution}, when power accuracy is required,
using (\ref{eq:spt_fixed}) is cumbersome since the number of relevant terms
varies as a function  of $Q$ and $\rho$.  To evaluate $\ln J(\nu,Q^2)$ we
therefore perform explicitly the Borel integral~(\ref{J_nu_Borel}). The
prescription we use to define the integral avoiding the renormalon
singularities as well as the technique used for the analytic integration were
explained in detail in~\cite{Thrust_distribution}.

To complete the perturbative calculation we need to match the resummed result
with the exact NLO calculation, which includes non-logarithmic contributions.
We do this using the so-called $\log R$ matching scheme~\cite{Catani:1993ua}. 
This amounts to replacing the first two terms in the expansion of 
$\ln R(Q^2,y)$ (where $y=t,\rho_H$) with the exact terms in the 
following way:
\begin{eqnarray*}
\left.\ln R(Q^2,y)\right\vert_{\rm PT} & = &
R_1(y)a(Q^2)+\left(R_2(y)-\frac{1}{2}R_1(y)^2\right)a^2(Q^2) \\ &&
+ \left.\ln R(Q^2,y)\right\vert_{\rm DGE}
- \left.\ln R(Q^2,y)\right\vert_{\rm DGE}^{\rm NLO} \; .
\end{eqnarray*}
Here $a=\alpha_{\overline{\mbox{\tiny MS}}}/\pi$, $R_1(y)$ and
$R_2(y)$ denote the exact LO and NLO coefficients respectively,
and $\left.\ln R(Q^2,y)\right\vert_{\rm DGE}^{\rm NLO}$ is the
expansion of the DGE resummed result to ${\cal O}(a^2)$. For
$R_1(y)$ we use the analytic results and for $R_2(y)$ we use
formulae that parametrize the coefficients extracted numerically 
from EVENT2~\cite{EVENT2}. Whereas the resummed DGE result is independent
of the renormalization scale\footnote{It should be stressed that independence of the renormalization scale applies within a given renormalization scheme -- the
scheme in which $\bar{\alpha}_s(k^2)$ is defined. Here the scheme is
fixed {\` a} la `t Hooft, so that the renormalization group equation has
only two non-vanishing coefficients. This arbitrary choice
is reflected in terms that are subleading in $\beta_0$ at the exponent, at
NNLL accuracy and beyond.}, the matched result has some scale dependence. 
Ignoring the renormalon contributions which are not logarithmically enhanced 
is justified a posteriori: 
in~\cite{Thrust_distribution} the residual scale dependence 
was found to be very small for the thrust distribution ($\lsim 1\%$ for $t<0.25$). We expect a similar dependence for the heavy-jet mass.

\subsection{Limitations of the approximation used}
\label{sec:applicability}

The perturbative DGE result presented in the previous sections was the basis
for the phenomenological analysis of the thrust distribution
in~\cite{Thrust_distribution} using (almost)  the entire range in $t$.
However, as will soon become clear, the range of applicability of the result
for the heavy-jet mass is more restricted. To understand why, and eventually
in what conditions a quantitative study of the heavy-jet mass can be
performed, we will now return to some of the approximations made in the
derivation and discuss their validity.

As any resummation, the DGE result captures specific features of the
distribution, which show up as large corrections, neglecting other,
sub-dominant features. To judge  the quality of the approximation, we must
know what we neglect. The first crucial comparison is, of  course, with
fixed-order results. Although the known fixed-order results (here LO and NLO)
are taken into account by matching, their comparison with the resummed result
is a valuable tool to identify the range of parameters where the
approximation is valid and what it may miss  at yet unavailable higher orders
(NNLO and beyond).

\begin{figure}[t]
\center \epsfig{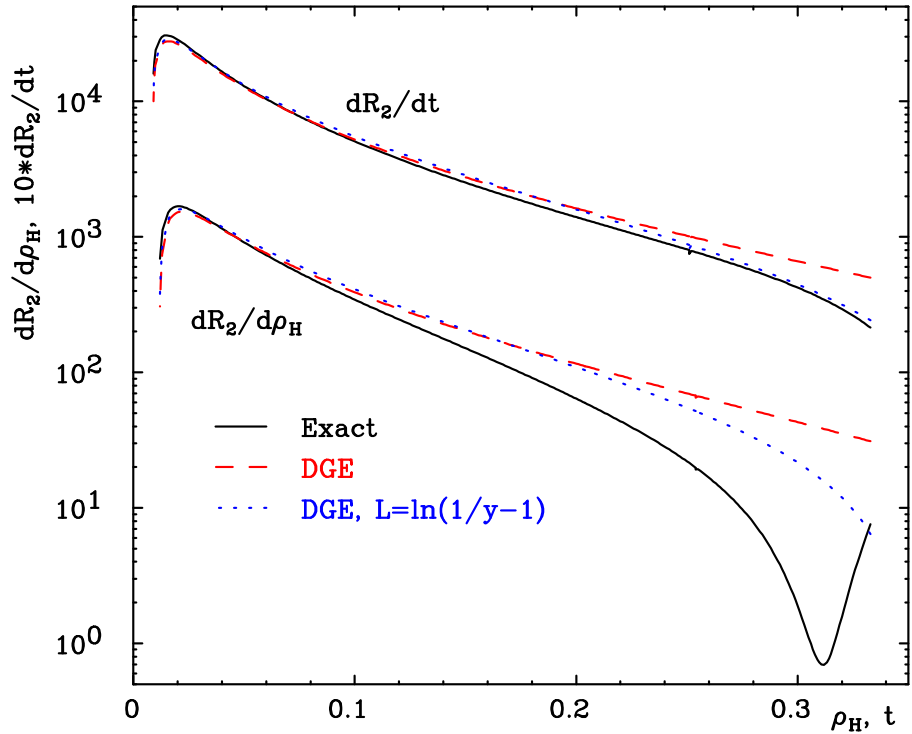}
\caption{The NLO coefficient for the thrust (upper set of curves)
and heavy-jet mass (lower set of curves) distributions obtained by
expanding the DGE result (dashed and dotted line for the
unmodified and modified log,~respectively) compared to the exact
result (solid line). } \label{fig:nlo}
\end{figure}

The comparison of the ${\cal O}(\alpha_s^2)$ term in the resummed result,
expanded as \[R(Q^2,y)=1+R_1(y)a(Q^2)+R_2(y)a^2(Q^2)+\ldots,\] with the exact
NLO term calculated by EVENT2 \cite{EVENT2}, is shown in Fig.~\ref{fig:nlo}.
The difference between the two increases for $y \gg y({\rm peak})$,
particularly because of the non-logarithmic terms. The approximation for large
values of the variable is much better in the case of the thrust compared to
that of the heavy-jet mass. For example, the ratio between $dR_2^{\rm
NLO}(t)/dt$ and $dR_2^{\rm DGE}(t)/dt$ is larger than $0.8$ for $t\leq 0.25$,
whereas for the heavy-jet mass the corresponding ratio reaches $0.8$ already
at $\rho_H=0.13$.
The figure also shows that for the thrust distribution a modification of the
logarithm to $L=\ln(1/t-1)$, as suggested in~\cite{Catani:1991kz}, gives a
better approximation for large $t$, whereas for intermediate $t$ ($\sim 0.1$)
the approximation with a modified log is worse. In the case of the heavy-jet
mass the effect of modifying the log is much smaller than the difference with
the exact result. We conclude that in the latter case modifying the log cannot
be considered as a reliable estimate of the uncertainty due to the missing
non-logarithmic higher-order terms. We will explain why this is so, below.

Having based our exponentiation on the large $\beta_0$ limit,
rather than on a fixed logarithmic accuracy, it is interesting to
know which logs are computed exactly. As before, we categorize the
logs as they appear in the exponent, $\ln J(\nu,Q^2)$,
consistently with~\cite{Catani:1993ua}: leading logs are ${\cal
O}(L^{n+1}\alpha_s^n)$, NLL are ${\cal O}(L^{n}\alpha_s^n)$, etc.
In addition we must distinguish, at the level of the exponent, between terms
that are leading in $\beta_0$ which are directly calculated
(section~2.1) and terms that are subleading in $\beta_0$,
e.g.~${\cal O}(C_F C_A L^{j}\alpha_s^2)$, which are introduced
through subleading terms in the splitting function or through the
(two-loop) evolution of the coupling.

Terms that are leading in $\beta_0$ at the exponent would have been exact if
not for non-inclusive contributions~\cite{Nason:1995hd}. Such contributions
first show up due to the emission of a soft gluon at large angle, that splits
into a quark and an antiquark which end up in opposite hemispheres with
respect to the thrust axis. The contribution of such an emission to the single
jet mass $\ln J(\nu,Q^2)$ is ${\cal O}(C_F \beta_0 L \alpha_s^2)$ i.e. NNLL.
However, both the thrust and the heavy-jet mass distributions receive only a
${\cal O}(C_F \beta_0 \alpha_s^2)$ correction, i.e. $\rm N^{(3)}LL$.  The
reason is that in the latter case, contrary to the former\footnote{We are
grateful to Gavin Salam for explaining this point.}, the observables
and their inclusive approximations are of the same order -- the order of the
soft gluon momentum. Consequently, the difference between the integrated cross
section and its inclusive approximation has support only in a small part of
the phase space. The ${\cal O}(C_F \beta_0 L \alpha_s^2)$ term obtained from
expanding the inclusive DGE result indeed coincides with the numerical value
extracted from EVENT2. This was checked in the case of the thrust
in~\cite{Thrust_distribution} (see the appendix). We have now made a similar 
check for the heavy-jet mass.

Terms that are subleading in $\beta_0$ at the exponent are
computed exactly to NLL accuracy for both the thrust (\ref{eq:t})
and the heavy-jet mass (\ref{eq:hjm}) distributions, thanks to the
use of the two-loop renormalization group equation and the gluon
bremsstrahlung effective charge. The latter includes the full
non-Abelian contribution to the singular part of the splitting
function~\cite{Catani:1991rr,Thrust_distribution}. As in the
Abelian case discussed above, the inclusive approximation has a
price. The effect of a large-angle soft gluon that splits into two
gluons which end up in opposite hemispheres is larger than in
the Abelian case, because the second branching contains an
additional soft singularity. Indeed, Dasgupta and Salam have
shown~\cite{Dasgupta:2001sh} that the single-jet mass $\ln
J(\nu,Q^2)$ receives a ${\cal O}(C_F C_A L^2 \alpha_s^2)$ 
contribution, i.e. NLL, from such a configuration. As in the Abelian
case, the corresponding contributions to the thrust and the heavy
jet mass are suppressed, and appear at NNLL accuracy or beyond.

This discussion teaches us that the logarithmically enhanced terms
are well under control for both the thrust and the heavy-jet mass
distributions. The key to understanding the difference between
these two cases, as reflected in Fig.~\ref{fig:nlo} is the
kinematic constraints on the jet masses, which manifest themselves
primarily through non-logarithmic corrections.

When a hard gluon is emitted at a moderate angle (the heavy-jet mass is not
very small) emission of further soft gluons is not controlled by the same
matrix element. Also the phase space is drastically modified.  This is not
taken into account in our approximate calculation,  where gluons are emitted
off a quark--antiquark dipole. Moreover, we explicitly assumed in the
derivation of~(\ref{eq:hjm}) that the masses of the two hemispheres are
independent. This assumption is consistent with the approximation of
independent emission we employ in the two-jet region. However, it is clearly
violated at larger $\rho_H$.

\begin{figure}[t]
\center \epsfig{file=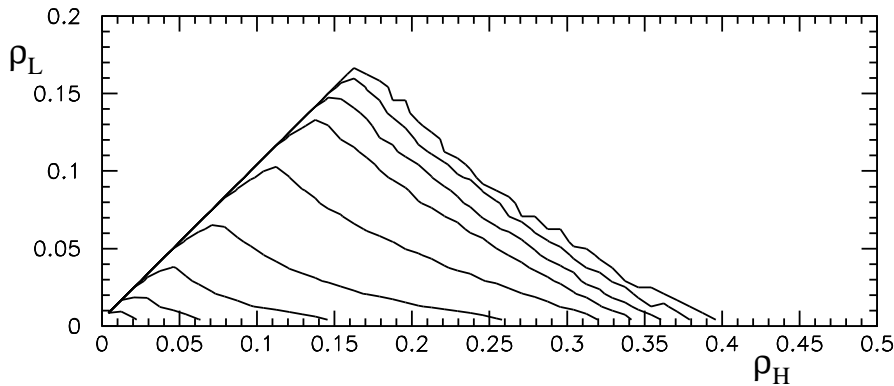,height=5.5cm} \caption{A
contour plot of the doubly differential light and heavy-jet mass
cross section based on 10 million events generated with {\sc Pythia} at
$Q={M_{\rm Z}}$ on the parton level, using only light quarks (uds). The
contours spacing is based on a logarithmic scale: the lowest
contour corresponds to two events and the others are spaced by a
relative factor of 5. This illustrates the available phase space
and, in particular, the constraints on $\rho_L$ for $\rho_H >
1/6$.} \label{fig:phase}
\end{figure}

Contrary to the approximation we use, in a realistic 
phase space, the two hemisphere masses are correlated.  The
easiest way to see this is to consider a slightly different
definition of the jet masses, where the hemispheres are defined by
minimizing the sum of the two jet masses $\rho_{\rm tot}\equiv
\rho_1+\rho_2$.  In this case it was shown~\cite{Clavelli:1981yh} 
that the largest possible value of $\rho_{\rm tot}$ is
$1/3$. This means that $\rho_L$ cannot
exceed $\min\left\{\rho_H, 1/3-\rho_H\right\}$, which is always
smaller than $1/6$.  
This constraint becomes quite stringent for
large values of $\rho_H$. It manifests itself through a Sudakov
form factor, which resums logarithms of the form $\ln
(1/3-\rho_H)$. These large perturbative corrections have nothing
to do with the $\ln \rho_H$ terms that are resummed in the two-jet
region. 

Turning to the jet mass definition based on the thrust axis, the
upper limit changes to $\rho_1+\rho_2 \leq 1/2$, but the value of
$\rho_H$ for which $\rho_L$ reaches its maximum is still $1/6$
(this value does not depend on the definition of the thrust axis,
since at this point the distribution of particles is completely
symmetric). The available phase space for $\rho_L$ as a function
of $\rho_H$ is illustrated in Fig.~\ref{fig:phase}. The figure
shows the doubly differential cross section based on 10 million
events generated with {\sc Pythia}~\cite{PYTHIA} for $Q={M_{\rm Z}}$ at
the parton level.

Ignoring the kinematic constraints in the calculation, we definitely
overestimate the cross section for $\rho_H > 1/6$. Since the cross section for
$\rho_H\geq 1/6$ is not negligible (it is of the order of $\sim 5\%$ 
at $Q={M_{\rm Z}}$), there
is also an impact on the cross section below $1/6$. This explains the
significant difference (Fig.~\ref{fig:nlo}) between the exact result and our
approximation, as $\rho_H$ increases.

The kinematic constraint on the light-jet mass for $\rho_H>1/6$ is much more
stringent than the general condition that the cross section vanishes at
$\rho_H=1/2$. Whereas the former condition dictates the behaviour of the exact
NLO coefficient in Fig.~\ref{fig:nlo}, it is the latter that is imposed by
modifying the log from $\ln 1/\rho_H$ to $\ln (1/\rho_H-1)$. This explains why
such a modification does not represent the actual effect of non-logarithmic
terms at large $\rho_H$.

The reason why the effect on the thrust distribution is smaller also becomes
clear: fixing $t$ at some moderate value does not put any stringent constraint
on the light-jet mass, nor on any other shape variable. When $t$ reaches $1/3$
such constraints appear. However, since the cross section for $t>1/3$ is small
(less than $1\%$
at $Q={M_{\rm Z}}$), 
the effect of ignoring the kinematic constraints on the
cross section below $1/3$ is insignificant.

It should also be emphasized that the correlation between the hemisphere
masses  due to the restriction on their sum is {\em negative}. This should be
contrasted with the dynamical  correlation discussed
in~\cite{Korchemsky:2000kp,Belitsky:2001ij},  which is due to radiation from
one hemisphere into the other (``non-inclusive''
corrections~\cite{Nason:1995hd}).  This correlation is  {\em
positive}~\cite{Gardi:2000yh,Korchemsky:2000kp,Belitsky:2001ij}. Recall that
the non-inclusive correction at the NLO level was shown~\cite{Gardi:2000yh}
(see table 2 there, and the explanation that follows) to be small in the case
of the average thrust, but it increases for higher moments $\langle t^2 \rangle$, $\langle t^3 \rangle$,
etc. Based on the definitions (\ref{eq:tdef}) and (\ref{eq:hdef}), both $t$ and
$\rho_H$ are smaller in the full calculation than in the inclusive one, in
configurations where a gluon decays into particles that end up in opposite
hemispheres. Therefore, in the full, non-inclusive calculation the thrust
distribution is somewhat flatter than in the inclusive one~\cite{Gardi:2000yh}.
The same holds for the $\rho_H$ distribution. Thus, ignoring the radiation from
one hemisphere into the other results in underestimating both the thrust and
the heavy-jet mass distributions at large $t$ and $\rho_H$, respectively. On
the other hand, Fig.~\ref{fig:nlo} shows that ignoring the kinematic
constraints as well as the radiation from one hemisphere into the other,
results in overestimating the cross section at large $\rho_H$, whereas the
effect on the thrust is smaller. We conclude that the dominant feature of the
$\rho_H$ distribution missed by the current approximation is due to the
kinematic constraints (non-logarithmic terms). 

Finally, the main conclusion from this analysis is that, for the
heavy-jet mass distribution, an approximation based only on the
two-jet limit~(\ref{eq:hjm}) breaks down already at $\rho_H=1/6$
and, in general, is not as accurate as the analogous calculation
of the thrust distribution~(\ref{eq:t}). Of course, the
approximation is improved by matching the resummed expression with
fixed order calculations, but in the absence of a NNLO result,
large uncertainties are unavoidable. Since the cross section is
peaked at rather small $\rho_H$, it is worth while to attempt a
quantitative study of the distribution in the two-jet region,
using an upper cut on the analysed data at $\rho_H = 1/6$ or
below. In performing such an analysis one must keep in mind possible
large corrections beyond the NLO.

\subsection{Non-perturbative corrections}
\label{sec:np}

Generally, the effect of hadronization on an event-shape observable
can be taken into account by power 
corrections~\cite{Average_thrust}--\cite{Salam:2001bd}. 
The presence of power corrections can be deduced from perturbation theory: 
the renormalon ambiguity which appears upon resumming the perturbative
expansion must be compensated on the non-perturbative level. This allows us to
detect, by perturbative tools, certain non-perturbative corrections. Assuming
that these corrections dominate, we can parametrize the hadronization
corrections based on the form of the renormalon ambiguity, using a small set of
parameters.

Since the DGE formulae for the thrust (\ref{eq:t}) and the heavy-jet mass
(\ref{eq:hjm}) distributions are both expressed in terms of the single-jet
mass distribution $\ln J(\nu,Q^2)$, it is natural to parametrize the power
corrections through this quantity. This immediately implies a relation between
the corrections to the two distributions, a relation that can be confronted
with data. In parametrizing the non-perturbative corrections through $\ln
J(\nu,Q^2)$, we will effectively regard 
(\ref{eq:t}) and (\ref{eq:hjm}) as non-perturbative relations. 
This means in particular that we will ignore correlations
between the hemispheres also on the non-perturbative level. This should be
contrasted with~\cite{Korchemsky:2000kp}, where such correlations were found
to be {\em essential} for the consistency between the heavy-jet mass 
distribution and that of the thrust.

As discussed in~\cite{Thrust_distribution}, the first thing we learn from the
perturbative formula for DGE~(\ref{J_nu_Borel}) is that ambiguities, and
therefore power corrections, appear in the {\em exponent}. Next, the renormalon
singularities of $\ln J(\nu,Q^2)$ in~(\ref{Borel_nu}) (taking into account
the factor $\sin \pi u/\pi u$ of Eq.~(\ref{J_nu_Borel})) appear at half
integers~$u=\frac12,\frac32, \frac52 \ldots$, and at $u=1,2$. The
singularities at $u=1, 2$ are due to collinear emission. They indicate power
corrections of the form~$\sim 1/(Q^2\rho)$ and~$\sim 1/(Q^4\rho^2)$, which can
be neglected in the region~$\rho Q/\gsim \bar{\Lambda}$. The half-integer
singularities are associated with the large-angle soft emission. They
indicate~\cite{Thrust_distribution} power corrections of the form~$\sim
1/(Q\rho)^n$, where $n$ is an {\em odd} integer. These power corrections are
dominant as~$\rho$ approaches~$\bar{\Lambda}/Q$, and they can be resummed into
a non-perturbative shape function of a single argument, as originally
suggested by Korchemsky and
Sterman~\hbox{\cite{Korchemsky:1995is}--\cite{Belitsky:2001ij}.} 

Possible modifications of the Borel singularities beyond the large $\beta_0$
limit were also examined in~\cite{Thrust_distribution}. It was shown that,
through the modification of the Borel poles into cuts, both large-angle, soft
gluon corrections of even powers and collinear corrections, whose magnitude is
controlled by the same parameters, are present.  The phenomenological
analysis below will follow~\cite{Thrust_distribution} and include all
these power corrections:  the large-angle, soft gluon corrections of {\em any}
power will be parametrized by a single argument shape function, as done
in~\cite{Korchemsky:1999kt,Korchemsky:2000kp},  and the leading power
corrections of  collinear origin will be introduced  even though
the analysis of~\cite{Thrust_distribution} showed that the effect of the 
latter is very small.

The integrated cross section of the single-jet mass is then given by,
\begin{eqnarray}
\label{PT_NP}
R_{\rho}(Q^2,\rho)
  & =  &
\int_C \frac{d\nu}{2\pi i\nu} \,
  \exp \left[\nu \rho + \ln J^{\rm PT}(\nu,Q^2) + \ln J^{\rm NP}(\nu,Q^2)  \right],
\end{eqnarray}
where the {\em sum} of the perturbative (PT) and non-perturbative (NP) 
contributions to the exponent is a physical quantity and thus independent of the
renormalon regularization prescription. Our prescription for the Borel
integration, which defines the perturbative part $\ln J^{\rm PT}(\nu,Q^2)$ of
Eq.~(\ref{J_nu_Borel}), was described in~\cite{Thrust_distribution}. 
Note that it is
different from the cutoff definition used
in~\cite{Korchemsky:1999kt,Korchemsky:2000kp}.

Thanks to the factorized form in $\nu$-space, Eq.~(\ref{PT_NP}) can be 
written as a convolution:
\begin{eqnarray}
\label{conv}
R_{\rho}(Q^2,\rho)  =
\int_0^\rho \,d\tilde{\rho}  \,
R^{\rm PT}_{\rho}(Q^2,\rho-\tilde{\rho}) \,
f_{\rm NP}(Q^2,\tilde{\rho})   \equiv
R^{\rm PT}_{\rho}(Q^2,\rho) \otimes f_{\rm NP}(Q^2,\rho)
\end{eqnarray}
where the non-perturbative correction is
\begin{eqnarray}
\label{f_NP}
f_{\rm NP}(Q^2,\rho)
  =
\int_C \frac{d\nu}{2\pi i}\,
  \exp \left[ \nu \rho+\ln J^{\rm NP}(\nu,Q^2)  \right].
\end{eqnarray}
Given that $f_{\rm NP}(Q^2,\rho)$ incorporates the non-perturbative corrections
to the single-jet mass distribution, and given the relations (\ref{eq:sjm}) and 
(\ref{eq:t}), i.e.
\begin{eqnarray*}
&&R_{t}(Q^2,t)=
R_{\rho}(Q^2,t)\otimes
R_{\rho}(Q^2,t) \\
&&R_{\rho_H}(Q^2,\rho_H)=
\left\{R_{\rho}(Q^2,t)\right\}^2,
\end{eqnarray*}
the corrections to the thrust and heavy-jet mass distributions can then
be expressed as:
\begin{eqnarray}
&&R_{t}(Q^2,t)
   =
R^{\rm PT}_{t}(Q^2,t)  \otimes
f_{\rm NP}(Q^2,t) \otimes f_{\rm NP}(Q^2,t)
 \nonumber \\
&&R_{\rho_H}(Q^2,\rho_H)
   =
\left\{
\left[R^{\rm PT}_{\rho_H}(Q^2,\rho_H)\right]^{1/2}
\otimes f_{\rm NP}(Q^2,\rho_H)   \right\}^2.
\label{eq:convth}
\end{eqnarray}

Next, to parametrize $f_{\rm NP}(Q^2,\rho)$, we use what we learned from the
renormalon structure of the exponent. Ignoring, for the time being, corrections of
collinear origin, which appear beyond the large $\beta_0$ limit, we start, as
in~\cite{Korchemsky:1999kt}, by writing
\begin{eqnarray}
\label{eq:moments}
 \ln J^{\rm NP}(\nu,Q^2) \,\longrightarrow\, \ln J^{\rm NP}(\nu\bar{\Lambda}/Q) =
  -\sum_{n=1}^{\infty}
  \lambda_n\frac{1}{n!}\left(\frac{\nu\bar{\Lambda}}{Q}\right)^n,
\end{eqnarray}
where we exhibited the dependence of the non-perturbative function $J^{\rm
NP}(\nu\bar{\Lambda}/Q)$ on a single argument. According to the renormalon
ambiguity pattern of~(\ref{Borel_nu}), we expect corrections of odd powers but
not  of even powers. Thus the relevant parameters are $\lambda_n$ with odd~$n$.
Following~\cite{Korchemsky:1999kt,Thrust_distribution} the set of parameters
$\lambda_n$ can be traded for a single-argument shape function,
\begin{eqnarray}
 J^{\rm NP}(\nu\bar{\Lambda}/Q)=\int_0^{\infty} d\zeta \,f\left(\zeta \right)
\,\exp\left({-\zeta \nu\bar{\Lambda}/Q}\right).
\end{eqnarray}
This implies \cite{Korchemsky:1999kt} that $\lambda_n$ are the central moments
of the shape function, namely,
\begin{eqnarray*}
\lambda_1
&=& \int_0^\infty  d\zeta \,\zeta\, f(\zeta ) \\
\lambda_2
&=& -\int_0^\infty d\zeta\,   \left(\zeta-\lambda_1\right)^2 \,f (\zeta ),
\end{eqnarray*}
etc.
For small $\nu$ (large $\rho$) it is sufficient to keep the leading term in
$\ln J^{\rm NP}(\nu\bar{\Lambda}/Q)$, which gives
\begin{eqnarray}
\label{eq:shift}
f(\zeta)     =  \delta \left(\zeta-\lambda_1\right) .
\end{eqnarray}
Thus, for $\rho \gg \bar{\Lambda}/Q$ the non-perturbative effects can be
approximated by a shift of the perturbative distribution, in agreement
with~\cite{Shape_function2,Dokshitzer:1997ew}. Based on the expectation that the~$\lambda_2$
correction is suppressed, the shift approximation may hold in a rather wide
range.  However, in the distribution peak region higher powers in
(\ref{eq:moments}) become relevant and a shape function is required. For this
function we will be using the form suggested in~\cite{Thrust_distribution}:
\begin{equation}
\label{eq:shape}
f(\zeta )  =
n_0\zeta^q(1+k_1\zeta+k_2\zeta^2)e^{-b_1\zeta-b_2\zeta^2},
\end{equation}
which is flexible enough for the second central moment $\lambda_2$ to
vanish, as predicted by the large $\beta_0$ ambiguity pattern, but we will
usually not impose this condition. Here $n_0$ is fixed by normalization,
$\int_0^\infty d \zeta f(\zeta)  =1$, and $q,k_1,k_2,b_1,b_2$ are free
parameters to be determined by the data.

Finally, the full non-perturbative function (\ref{f_NP}), which we convolute with the perturbative distribution~(\ref{conv}), is
\begin{eqnarray}
\label{f_NP_conv}
f_{\rm NP}(Q^2,\rho)
=\left[\frac{Q}{\bar{\Lambda}}\,f\left(\frac{Q\rho}{\bar{\Lambda}}\right)\right]
\,
\otimes\,\left[\delta(\rho)-\lambda_2\left[\frac{{\bar{\Lambda}}^2}
{Q^2\rho}\right]_{+}\!\!\!\!\otimes\delta^{(1)}(\rho)\right],
\end{eqnarray}
where the first brackets contain the main, large-angle, soft gluon correction
(\ref{eq:shape}) and the second brackets incorporate the leading 
collinear correction (see~\cite{Thrust_distribution}).

Another aspect that should be taken into account is that the perturbative
calculation of the distributions is for massless partons, whereas  the detected
hadrons are massive. The effect of finite hadron masses was studied in detail
for average values of event-shape variables by Salam and
Wicke~\cite{Salam:2001bd}. They define two types of corrections:  universal,
which arise from the reshuffling of momenta when forming massive hadrons from
massless partons, and non-universal, which depends on the way the masses are
treated in the definition of the variables. Based on kinematic arguments the
mass effects were found to be of order $\bar{\Lambda}/Q$, whereas a more detailed QCD
analysis, using local parton -- hadron duality, indicated that both corrections
scale as $(\ln{Q}/\bar{\Lambda})^A\bar{\Lambda}/Q$, where $A=C_A/\beta_0\sim1.6$. The
origin of this logarithmic dependence~\cite{Salam:2001bd} is the fact that the
mass effects associated with transforming kinetic energy into hadron masses
grow with particle multiplicity, which increases logarithmically with the
energy.

The comparison between the hadronization corrections to the thrust and those to the heavy-jet mass is particularly problematic because the thrust is defined using
three-momenta, whereas the heavy-jet mass using four-momenta. According
to~\cite{Salam:2001bd}, this implies that the two variables will have
different sensitivity to the hadron masses, i.e. different non-universal mass
effects. This problem can be by-passed~\cite{Salam:2001bd}, modifying
the definition for the variables. In addition to the actual definition, the ``hadron level'' at which the measurement is performed makes a difference: depending on the experimental set-up, different particles may effectively be stable.

The non-perturbative parameters used to describe the
hadronization corrections may be defined in different ways. In particular, they 
depend on the specific definition of the variables, on the ``hadron
level'' used for the measurement, and on the way non-perturbative corrections 
are separated from the perturbative sum. 
Moreover, there is no unique way to distinguish between the hadronization process (the ``standard'' power corrections) and the formation of {\em massive} particles (the ``mass effects''). Therefore, contrary to $\alpha_s$, where the dependence on the particular procedure must be interpreted as some ``theoretical uncertainty'', the numerical values of the non-perturbative parameters are meaningful only within a given procedure. From this point of view it is satisfactory to have one procedure by which non-perturbative parameters to different event-shape observables can be compared.

In this paper we will consider the following procedures~\cite{Salam:2001bd} to define the variables, given that the detected particles are massive. We shall refer to these procedures as hadron mass schemes (HMS):
\begin{itemize}
\item {\bf Massive scheme (M)}: Use the massless definitions as given in
Eqs.~(\ref{eq:tdef})--(\ref{eq:ldef}),
ignoring the fact that the detected hadrons are massive. This
is the scheme that is normally used for the published data.
\item {\bf P scheme}: Use the measured three-momenta $\vec{p}_i$ as they are
and assign $E_i=|\vec{p}_i|$ for each particle. This way three-momentum is
conserved but not energy.
\item {\bf E scheme}: Use the measured energies $E_i$ as they are but rescale
the three-momenta with a factor $E_i/|\vec{p}_i|$. This way energy is conserved
but not three-momentum.
\item {\bf Decay scheme (D)}: Let all particles (including protons)
decay to massless particles. In the Monte Carlo this is done by 
 activating the decays of all particles
that are normally considered stable ($\mu^\pm$, $\pi^\pm$, $K^\pm$, $K^0_L$) .
In addition, we let the protons (and neutrons for simplicity) decay
semileptonically according to $p \to m e \nu$, where $m$ is massless. This
yields a final state consisting entirely of massless particles (the electron
mass can safely be neglected).
This way both energy and momentum are conserved,
but on the other hand the additional decays are not due to the strong 
interaction.
\end{itemize}
For the thrust and heavy-jet mass the non-universal mass effects that are
present in the conventionally used M scheme are diminished (or become common to
the two variables) in the decay, P and E schemes~\cite{Salam:2001bd}. The
decay scheme, in which energy and momentum conservation hold and all
``measured'' particles are strictly massless, seems the most attractive for
comparison with perturbative calculations. This will therefore be our preferred scheme. This choice is natural also according to the picture that emerges 
from the {\sc Ariadne}~\cite{Lonnblad:1992tz} Monte Carlo based study 
in~\cite{Salam:2001bd} (see Fig.~8 there): the
logarithmic dependence of the power correction coefficient is the smallest in
the decay level with respect to other hadronic levels.

To transform the data from one HMS to another we use the {\sc Pythia} Monte
Carlo model (version 6.158).  We have verified that the model describes well 
the thrust and the heavy-jet mass  distributions in the entire energy range and
in a wide range of the shape variables.  
Comparison with thrust data gives a~$\chi^2$
of~154 for 225 points for $14 \leq Q \leq 189$~GeV and $0.01\, {M_{\rm Z}}/Q < t
< 0.30$. Comparison with the heavy-jet mass data gives a $\chi^2$ of 129 for 209
points in the same ranges.
  
Using {\sc Pythia} we calculate transfer matrices
$T$ such that the cross section in bin~$i$ in scheme~$A$ is given
by $\sigma^A_i=\sum_jT^{A}_{ij}\sigma_j$, where $\sigma_j$ is the
cross section for bin $j$ in the original HMS. The matrices $T$
are calculated for each HMS, energy and experiment separately. For
the data at ${M_{\rm Z}}$ we use $10^7$ events to calculate the
matrices, whereas for the other energies we use $10^6$ events. This
way the statistical error in the calculation of $T$ can be
neglected with respect to the statistical error in the data. It should
be emphasized that this procedure is very similar to what is
normally used to make corrections to the data (e.g. those due to imperfect detector efficiency). To estimate the systematic error the calculation of $T$
one could be repeated using a different Monte Carlo
model such as {\sc Herwig}~\cite{HERWIG}. However, to do it properly
one should use the same Monte Carlo for the complete analysis
chain used for extracting the data.


\section{Data analysis}

In this section we compare the calculations of the thrust and the
heavy-jet mass distributions to the world data and perform fits to
extract $\alpha_s$ and the parameters that control the
hadronization corrections. A similar analysis for the thrust data
was performed in~\cite{Thrust_distribution}. Here, our main
motivation is the comparison between the thrust and the heavy-jet
mass. To make a fully quantitative comparison we must repeat the
thrust analysis taking into account several {\em experimental}
aspects that were not fully considered before. This includes the following issues:
\begin{description}
\item{\,\, a)\,\,\,\,} {\bf Using data where the measurement is corrected 
to be a measurement of all particles.}
Most experiments base their analysis on charged tracks and then
correct to include all particles, using a Monte Carlo model such
as {\sc Pythia}. However, some experiments have chosen to present data
for charged particles only. While this procedure gives better
control of the systematic error, the measured distribution does
not correspond to the distribution calculated
perturbatively. Thus, we want to use data that have been
corrected to represent all particles. Out of the data sets we used
in our earlier analysis the important ones, which are for
charged particles only, are the ALEPH and DELPHI data sets at 
$M_{\rm Z}$. The difference in the ``ideal measurement'' used by different
experiments to correct
the data explains the discrepancy we found~\cite{Thrust_distribution}
between the different
data sets at this energy. In order to be able to use data from all
experiments, especially at ${M_{\rm Z}}$, we apply the method
described in section~\ref{sec:np} to transform the charged
particle data to be for all particles. We have not tried to estimate 
the systematic errors in this procedure.
\item{\,\, b)\,\,\,\,} {\bf Using a proper HMS, common to the two variables.}
The published data for the thrust and the heavy-jet mass is based
on the definitions~(\ref{eq:tdef}) and~(\ref{eq:hdef}),
respectively. While for the former the definition coincides with the P scheme,
the latter is strictly a ``massive scheme''. 
Following~\cite{Salam:2001bd} and our
discussion in section~\ref{sec:np}, comparison of the hadronization
corrections of the two distributions requires to use a common,
non-massive scheme. Our main analysis will be made in the decay scheme.
However, we will also compare the different HMS and verify that
$\alpha_s$ extracted from the fits does not change (recall that
a strong dependence was found in~\cite{Salam:2001bd}).
\item{\,\, c)\,\,\,\,} {\bf Estimating the effect of heavy primary quarks.}
The published data are based on all events, including those where heavy
quarks are produced, while the calculation always refers to 5
light flavours. The contribution of the quark mass to the variable, as well as the difference in the soft-gluon radiation
from heavy quarks compared to that from light quarks can be a source of error. In
order to estimate this error, we use {\sc Pythia} to calculate the
distribution based on all (udscb) or only light (uds) primary
quarks and then use the ratio $R_i$ of the two distributions to
transform the data bin by bin according to,
\begin{eqnarray}
\label{eq:uds_transf} \sigma^{\rm uds}_i=R_{i}\sigma^{\rm
udscb}_i.
\end{eqnarray}
Similarly to the HMS transformations, the calculations of the
$R_{i}$ are done for each HMS, energy and experiment separately,
taking into account the different binnings of the data.
\end{description}

Some comments are due concerning the data we use. As
in~\cite{Average_thrust,Thrust_distribution}, 
the data are in the range 14 to 189 GeV.
When several data sets are available from the same experiment and
energy, we use the latest one. We have removed some data sets with
large errors, and restricted ourselves to published data, where the
details on the Monte Carlo corrections (in particular, on whether the
data correspond to all particles or to charged particles only)
are available. The data used for the fits are summarized in
Table~\ref{chi_sq_exp} below. When available separately, the
systematic errors have been added in quadrature to the statistical
ones. When fitting the data we have calculated the integrated 
cross section for each bin
included in the fit. However, in the plots showing the data we
have followed the customary practice of plotting the data points
at the bin centres, even though this may be somewhat misleading
for large bin widths.

Given the theoretical uncertainties for the heavy-jet mass
distribution we start with a separate analysis of the thrust data,
then continue with a separate analysis of the heavy-jet mass data,
and finally analyse the heavy-jet mass data using the results of
the thrust analysis as input. This makes it possible to test the
DGE method and the assumptions we make on the hadronization
corrections.

\subsection{Thrust distribution}

We present here the analysis of the thrust
distribution. Similarly to~\cite{Thrust_distribution} we begin by
performing fits based on a shift of the DGE perturbative distribution
in a limited range~\cite{Dokshitzer:1997ew}, and then turn to 
shape-function-based
fits~\cite{Korchemsky:1999kt,Korchemsky:2000kp} in
a wider range. As explained above, we will also study the HMS dependence
and the effect of heavy quarks as well as the impact of using data for 
all particles instead of charged particles only.

\subsubsection{Fits with a shift}

The analysis is based on fitting $\alpha_s$ and a shift
$\lambda_1\bar{\Lambda}/Q$ of the perturbative distribution according to
(\ref{eq:shift}). This applies to $t$ values far enough above the peak region,
$t\gg \bar{\Lambda}/Q$.

\begin{figure}[t]
\center \epsfig{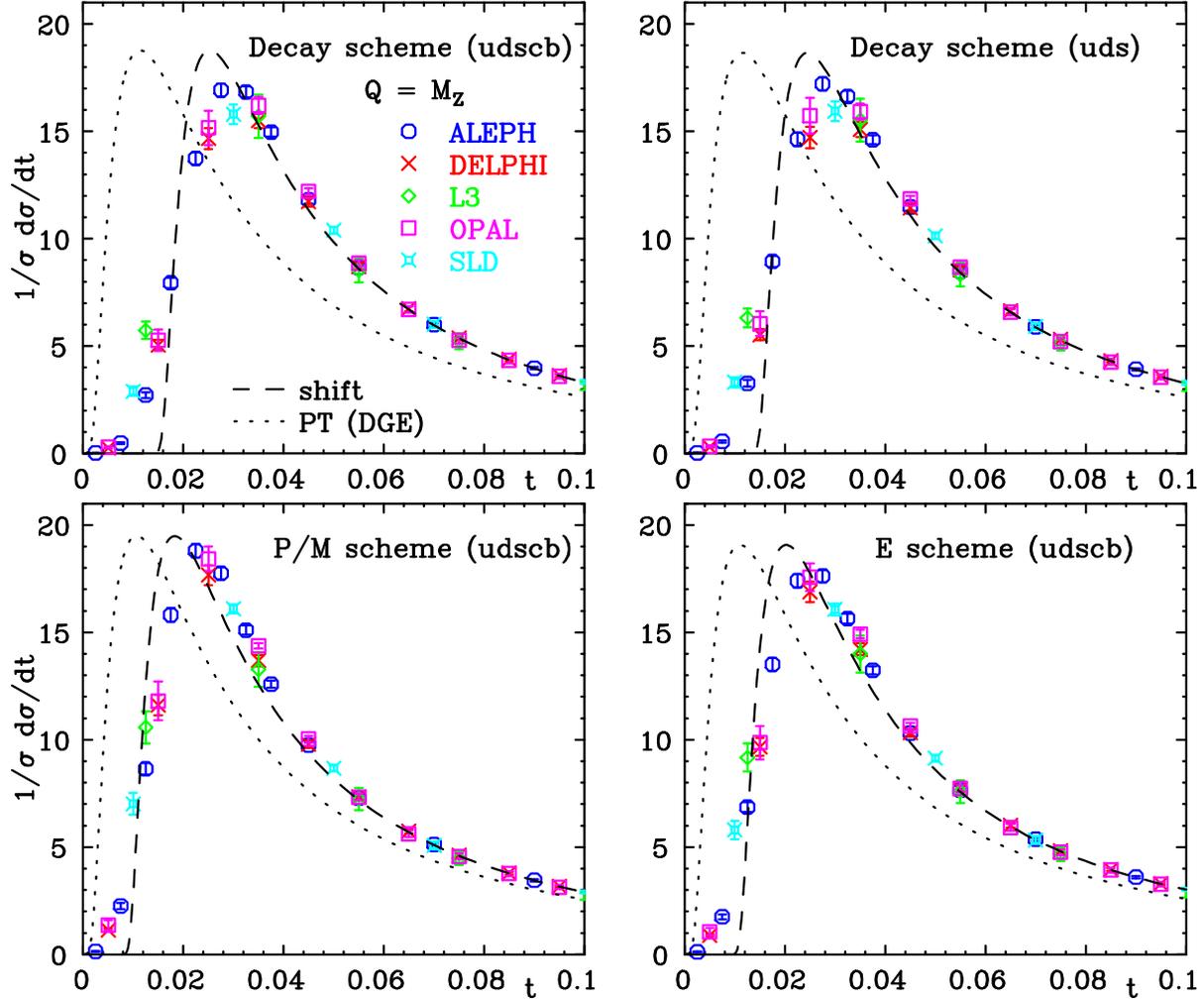}
\caption{Thrust data at ${M_{\rm Z}}$ in the peak region together
with shift-based fits using the range $0.05\,{M_{\rm Z}}/Q < t<
0.30$ (dashed line) in different HMS. The perturbative calculation
(DGE) is also shown (dotted line).} 
\label{fig:thrust_scheme_mz}
\end{figure}

We start by investigating the dependence on the HMS and the effect
of heavy primary quarks. Figure~\ref{fig:thrust_scheme_mz} shows the
fit results together with the data at ${M_{\rm Z}}$ in the different
HMS. We see that both the position and the height of the peak
vary. Note also the agreement between the different data sets in
the peak region in a given HMS (cf. Fig.~9
in~\cite{Thrust_distribution}) when the ALEPH and
DELPHI data have been corrected from charged particles only to all particles.

The results of fitting $\alpha_s$ and $\lambda_1$ to the data are
summarized\footnote{When comparing with~\cite{Thrust_distribution},
the reader should keep in mind that here the non-perturbative
corrections correspond to the single-jet mass $\ln J(\nu,Q^2)$
in (\ref{J_nu_Borel}) rather than to the exponent of the thrust,
$2\ln J(\nu,Q^2)$. Thus, the moments $\lambda_1$, $\lambda_2$,
etc., are smaller by a factor of $2$.} in Table~\ref{tab:scheme}.
The global impact of ignoring the difference 
between heavy and light primary quarks on the fit is very small and it will be 
neglected in most of the following analysis. This does not imply, 
of course, that the effect is small at low energies.  It is quite 
clear that it is significant at $14$~GeV,
but the weight of these data in the fit is small.
The table also shows that different HMS yield very similar
values of $\alpha_s$ (the differences are less than~$1.5\%$),
whereas~$\lambda_1$ changes by a factor of $\sim 2$.
\begin{table}[t]
\caption{ Fits of $\alpha_s$ and $\lambda_1$ to the thrust data with a lower
cut $t_{\min}=0.05\,{M_{\rm Z}}/Q$ and an upper cut $t_{\max}=0.30$
in different HMS.}
\begin{center}
\begin{tabular}{lcccc}
\tableline
 HMS & $\alpha_{\overline{\mbox{\tiny MS}}}({M_{\rm Z}})$ & $\lambda_1\bar{\Lambda}$ [GeV]& $\chi^2/{\rm dof} $ &  Points  \\
 M/P (udscb) &  0.1072 $\pm$ 0.0004   &  0.35  $\pm$  0.02  & 0.58  & 159 \\
 M/P (uds)   &  0.1074 $\pm$ 0.0005   &  0.31  $\pm$  0.02  & 0.58  & 159 \\
 E (udscb)   &  0.1080 $\pm$ 0.0004   &  0.42  $\pm$  0.02  & 0.51  & 159 \\
 E (uds)     &  0.1082 $\pm$ 0.0004   &  0.38  $\pm$  0.02  & 0.51  & 159 \\
 D (udscb)   &  0.1086 $\pm$ 0.0004   &  0.63  $\pm$  0.02  & 0.43  & 159 \\
 D (uds)     &  0.1088 $\pm$ 0.0004   &  0.59  $\pm$  0.02  & 0.45  & 159 \\
\tableline
\end{tabular}
\end{center}
\label{tab:scheme}
\end{table}
The stability of $\alpha_s$ is very important: a physical
parameter must not depend on arbitrary choices such as the
particular definition of the variable or the ``hadron level''.
Note that stability was not reached in~\cite{Salam:2001bd}, when
extracting $\alpha_s$ from mean values of event-shape variables.
Analysing the distribution is advantageous, since the dependence on
the variable, in addition to the energy, constrains $\alpha_s$
further. We have explicitly checked, performing a NLL
fit\footnote{The result in the decay scheme for all primary quarks
is $\alpha_{\overline{\mbox{\tiny MS}}}({M_{\rm Z}})  = 0.1217  \pm
0.0006$ and $\lambda_1\bar{\Lambda}= 0.69  \pm  0.02$ GeV with
$\chi^2/{\rm dof} =0.42$.} (with $\log R$ matching) to the
distribution that the results show the same stability as in the
DGE case. From the table it is also clear that the decay scheme
gives the best fits. We will mainly use this HMS in the following.

Next, we consider the dependence on the lower cut of the fit,
$t_{\min}$. The results are summarized in Table~\ref{tab:tmin}.
Note that for $t_{\min}\geq0.05\,{M_{\rm Z}}/Q$ the results are
stable: all fits for larger $t_{\min}$ agree within errors with
the result for $t_{\min}=0.05\,{M_{\rm Z}}/Q$.
\begin{table}[t]
\caption{Fits of $\alpha_s$ and $\lambda_1$ to the thrust data in the decay
scheme (udscb) as a function of the lower cut
$t_{\min}=t_{\min}({M_{\rm Z}})\,{M_{\rm Z}}/Q$ with the upper cut
fixed as $t_{\max}=0.30$.}
\begin{center}
\begin{tabular}{ccccc}
\tableline
$t_{\min}({M_{\rm Z}})$  & $\alpha_{\overline{\mbox{\tiny MS}}}({M_{\rm Z}})$ & $\lambda_1\bar{\Lambda}$ [GeV]& $\chi^2/{\rm dof} $ &  Points  \\
 0.03       &  0.1075 $\pm$ 0.0002    &  0.69  $\pm$  0.01  & 0.55  & 190 \\
 0.04       &  0.1075 $\pm$ 0.0003    &  0.69  $\pm$  0.01  & 0.54  & 173 \\
 0.05       &  0.1086 $\pm$ 0.0004    &  0.63  $\pm$  0.02  & 0.43  & 159 \\
 0.06       &  0.1091 $\pm$ 0.0005    &  0.60  $\pm$  0.03  & 0.42  & 147 \\
 0.07       &  0.1091 $\pm$ 0.0007    &  0.60  $\pm$  0.04  & 0.45  & 135 \\
 0.08       &  0.1090 $\pm$ 0.0007    &  0.61  $\pm$  0.04  & 0.43  & 124 \\
 0.09       &  0.1086 $\pm$ 0.0009    &  0.64  $\pm$  0.06  & 0.45  & 111 \\
 0.10       &  0.1084 $\pm$ 0.0011    &  0.65  $\pm$  0.07  & 0.48  & 104 \\
\tableline
\end{tabular}
\end{center}
\label{tab:tmin}
\end{table}
This means that there is essentially no dependence on the lower cut (it is less
than $0.5\%$), provided that the cut is far enough above the peak region.

We have checked that varying the upper cut $t_{\max}$ between
$0.15$ and $0.35$, while keeping the lower cut $t_{\min}=0.05\, 
M_{\rm Z}/Q$ fixed, the results for $\alpha_{\overline{\mbox{\tiny
MS}}}({M_{\rm Z}})$ and $\lambda_1$ agree, within errors, with the
results obtained for $t_{\max}=0.30$. Thus, the dependence on the
upper cut is also negligible (less than $0.5\%$).

To demonstrate the importance of using data for all particles
rather than for charged particles only, we made a separate fit to
the ALEPH and DELPHI data at ${M_{\rm Z}}$ in the P scheme. The
result changes from 
$\alpha_{\overline{\mbox{\tiny MS}}}({M_{\rm Z}})  = 0.1079  \pm
0.0005$ and $\lambda_1\bar{\Lambda}= 0.33  \pm  0.03$ GeV with
$\chi^2/{\rm dof} =0.45$ for all particles, to
$\alpha_{\overline{\mbox{\tiny MS}}}({\rm
M_Z})  = 0.1091 \pm 0.0005$ and $\lambda_1\bar{\Lambda}= 0.27  \pm
0.03$ GeV with $\chi^2/{\rm dof} =0.62$ for charged particles.
Thus, the difference
in the fit results is small, but not negligible. However, there is
a significant difference in the quality of the fit: in addition to
the lower $\chi^2/{\rm dof}$, the sensitivity to varying the cuts
on the fitting range is much lower. This concerns in particular the
lower cut -- compare Table~\ref{tab:tmin} above with Table 2
in~\cite{Thrust_distribution}.

\begin{figure}[t]
\center \epsfig{file=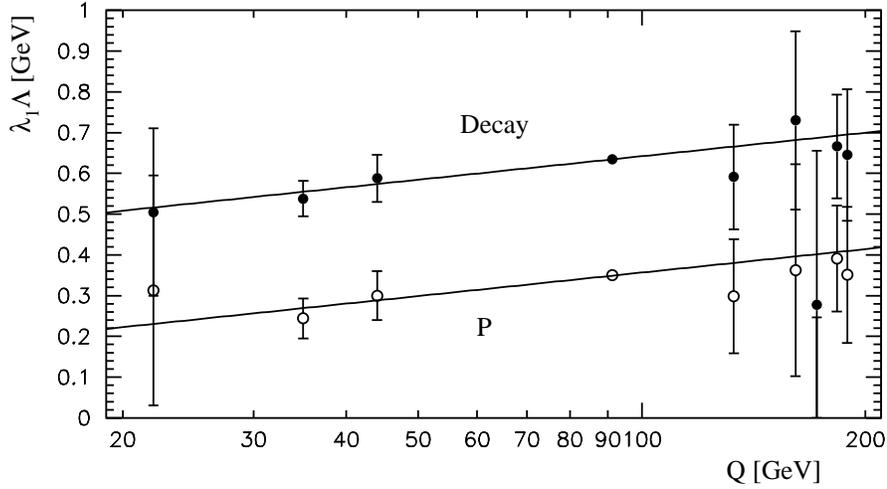,width=12cm} \caption{The
$Q$-dependence of the shift $\lambda_1$ for a fixed $\alpha_s$ in
the decay and P schemes. The lines are fits of the form
$\lambda_1\bar{\Lambda} =p_1+ p_2 \ln ({Q}/{\bar{\Lambda}})$.} 
\label{fig:qdep}
\end{figure}

Finally, knowing that beyond the large-$\beta_0$ limit power corrections are
modified by logarithmic $Q$ dependence~\cite{Thrust_distribution}, it 
is interesting to look for logarithmic $Q$ dependence of $\lambda_1$ by 
performing separate fits at each energy.
Additional motivation for such investigation is provided 
by the finding of~\cite{Salam:2001bd} that the scale dependence of 
hadron-mass related non-perturbative effects is enhanced by a factor of
$\ln^A({Q}/{\bar{\Lambda}})$, where $A=1.565$.
Fixing~$\alpha_s$ to the value obtained by fitting all energies,
we fitted $\lambda_1$ separately for each energy. The results of
these fits in the decay and P schemes are shown
in~Fig.~\ref{fig:qdep}, along with a fitted line of the form
$\lambda_1\bar{\Lambda} =p_1+ p_2\ln({Q}/{\bar{\Lambda}})$. The best-fit 
values of $p_1$ and $p_2$ are
\begin{eqnarray}
{\rm D \; (udscb): } &&
\begin{tabular}{l}
$p_1 = 0.13  \pm 0.11$  GeV  \\
$p_2 =  0.08  \pm  0.02$ GeV,
\end{tabular}
\end{eqnarray}
\begin{eqnarray}
{\rm P \; (udscb): } &&
\begin{tabular}{l}
$p_1  = -0.16  \pm 0.12$  GeV  \\
$p_2  =  0.08  \pm  0.02$ GeV.
\end{tabular}
\end{eqnarray}
The results show some trend of
increase in $\lambda_1$, but the power of the logarithmic dependence 
cannot be extracted from the data. The difference
between the decay and P schemes appears to be large, but
independent of $Q$. In the following we will ignore the
logarithmic $Q$ dependence and, instead, concentrate on fits with a
shape function and the determination of $\lambda_2$.

To summarize, the best fit values of $\alpha_s$ and $\lambda_1$
when using the decay scheme with all primary quarks, are
\begin{eqnarray}
\label{eq:thrust_shiftres}
{\rm D \; (udscb): } &&
\begin{tabular}{l}
$\alpha_{\overline{\mbox{\tiny MS}}}({M_{\rm Z}})  = 0.1086  \pm 0.0004$  \\
$\lambda_1\bar{\Lambda}= 0.63  \pm  0.02$ GeV
\end{tabular}
\end{eqnarray}
with $\chi^2/{\rm dof} =0.43$.
These values were found to be stable under variations of the lower and
upper cuts of the fitting range and to have very small uncertainty due
to heavy primary quarks. The uncertainty in the determination of
$\alpha_s$ based on varying the HMS was found to be~$\sim 2\%$.
The shift $\lambda_1$ can be a meaningful parameter only when the 
HMS has been fixed, in
which case it is determined up to $5$ -- $10$\% depending on the fitting range and the effect of heavy primary quarks. In the different schemes we considered, $\lambda_1$
changes by up to a factor $\sim 2$, giving an indication on the significance of hadron mass effects.

\subsubsection{Fits with a shape function}

Fitting with a shape function has the advantage that the peak
region can be included in the analysis and the structure of
subleading power corrections can be detected. While the
motivation is to extend the fitting range as much as possible, 
the region  $t \simeq \bar{\Lambda}/Q$, where all powers are equally
important, must be avoided. Similarly to~\cite{Thrust_distribution},
these considerations lead us to choose $t_{\min} =0.01\,
M_{\rm Z}/Q\simeq 4\bar{\Lambda}/Q$. The following fits are made to
data in the decay scheme (with all primary quarks).

For the functional form of the shape function~(\ref{eq:shape}), we
use two different possibilities for the fall-off at large $\zeta$:
exponential and Gaussian. The results for an exponential fall-off
are:
\begin{eqnarray}
\label{eq:thrust_shaperes}
{\rm D \; (udscb): } &&
\begin{tabular}{ll}
$\alpha_{\overline{\mbox{\tiny MS}}}({M_{\rm Z}}) = 0.1090 \pm 0.0004$&  \\
$\lambda_1\bar{\Lambda} = 0.605\pm 0.013 \; \; {\rm GeV}$ &
$\lambda_2\bar{\Lambda}^2 = 0.002\pm 0.024 \; \; {\rm GeV}^2$ \\
$b_1 = 0.61 \pm  0.10$  &$b_2 \equiv 0 \; \; {\rm (fixed)}$ \\
$q = 1.55  \pm 0.11 $&
\end{tabular}
\end{eqnarray}
with $\chi^2/{\rm dof}=0.53 $ for 225  points, and those for a Gaussian
fall-off:
\begin{eqnarray}
{\rm D \; (udscb): } &&
\begin{tabular}{ll}
$\alpha_{\overline{\mbox{\tiny MS}}}({M_{\rm Z}}) = 0.1085 \pm 0.0004$&  \\
$\lambda_1\bar{\Lambda} = 0.625 \pm 0.011 \; \; {\rm GeV}$ &
$\lambda_2\bar{\Lambda}^2 = -0.049\pm 0.018 \; \; {\rm GeV}^2$ \\
$b_1  \equiv 0 \; \; {\rm (fixed)}$  &$b_2 = 0.028 \pm  0.004$ \\
$q = 0.77  \pm 0.08 $&
\end{tabular}
\end{eqnarray}
with $\chi^2/{\rm dof}=0.58 $ for 225  points. Both results are in
good agreement with the shift-based fits for $t_{\min} \geq 0.05\,
{M_{\rm Z}}/Q$~(\ref{eq:thrust_shiftres}). As was the case in our
earlier analysis, the best fit turns out to have an exponential
fall-off. The Gaussian fall-off gives a slightly worse~$\chi^2$. 
In addition,
the softer, exponential fall-off is closer to the
asymptotic behaviour found in~\cite{Belitsky:2001ij} in a
calculation of the shape function using a perturbative model.

The sensitivity to the functional form gives a good indication of
the difficulty in determining $\lambda_2$ and higher moments of
the shape function from the data in the absence of more definite theoretical
input. The results of the exponential fall-off fit, which is the
best fit, are in agreement with the large $\beta_0$ renormalon ambiguity
pattern, which suggests that $\lambda_2=0$, but this is not true for the
Gaussian fall-off. Nevertheless, in both cases $\lambda_2 \ll
\lambda_1^2$. Thus, if $\lambda_1\bar{\Lambda}$ is used as a
reference scale for the non-perturbative parameters, then
$\lambda_2$ is always small. One should also keep in mind that the
value of $\lambda_2$ is strongly correlated to the values of
$\alpha_s$ and $\lambda_1$. For example, fixing the values of
$\alpha_s$ and $\lambda_1$ according to the fit with an exponential
fall-off when fitting with a Gaussian fall-off the result is
$\lambda_2\bar{\Lambda}^2 = -0.020  \pm 0.009 $ GeV$^2$, which
agrees, within errors, with the result for a exponential fall-off.

We have also verified that the result of the shape-function-based fit  is {\em not
sensitive} to the fitting range. This is true  for $\alpha_s$ as well as for the
non-perturbative parameters $\lambda_1$  and $\lambda_2$. In particular, the dependence
on the lower cut $t_{\min}$ is summarized in Table~\ref{tab:tminsf}. It is interesting
to note that the fact that $\lambda_2$ is small can be deduced even if the region to the left of the distribution peak is excluded from the fit. 
\begin{table}[t]
\caption{ 
Fits of $\alpha_{\overline{\mbox{\tiny MS}}}(M_{\rm Z})$ and shape function 
with exponential fall-off to the thrust data in the decay scheme (udscb)
as a function of the
lower cut $t_{\min}=t_{\min}(M_{\rm Z})M_{\rm Z}/Q$ with the upper cut
$t_{\max}=0.30$
fixed.}
\begin{center}
\begin{tabular}{cccccc} 
\tableline
$t_{\min}(M_{\rm Z})$ & $\alpha_{\overline{\mbox{\tiny MS}}}({M_{\rm Z}})$  & $\lambda_1\bar{\Lambda}$ [GeV]& $\lambda_2\bar{\Lambda}^2$ [GeV$^2$]& $\chi^2/{\rm dof} $ &  Points  \\  
 0.01           & 0.1090$\pm$ 0.0004  &  0.605  $\pm$  0.013  & 0.002   $\pm$ 0.024   & 0.53  &  225 \\
 0.02           & 0.1088$\pm$ 0.0004  &  0.619  $\pm$  0.015  & -0.004   $\pm$ 0.023   & 0.48  &  209 \\
 0.03           & 0.1089$\pm$ 0.0004  &  0.618  $\pm$  0.016  & -0.001   $\pm$ 0.023   & 0.47  &  190 \\
 0.04           & 0.1089$\pm$ 0.0006  &  0.623  $\pm$  0.040  & 0.015   $\pm$ 0.093   & 0.42  & 173  \\
\tableline
\end{tabular}
\end{center}
\label{tab:tminsf}
\end{table}

\begin{figure}[t]
\center \epsfig{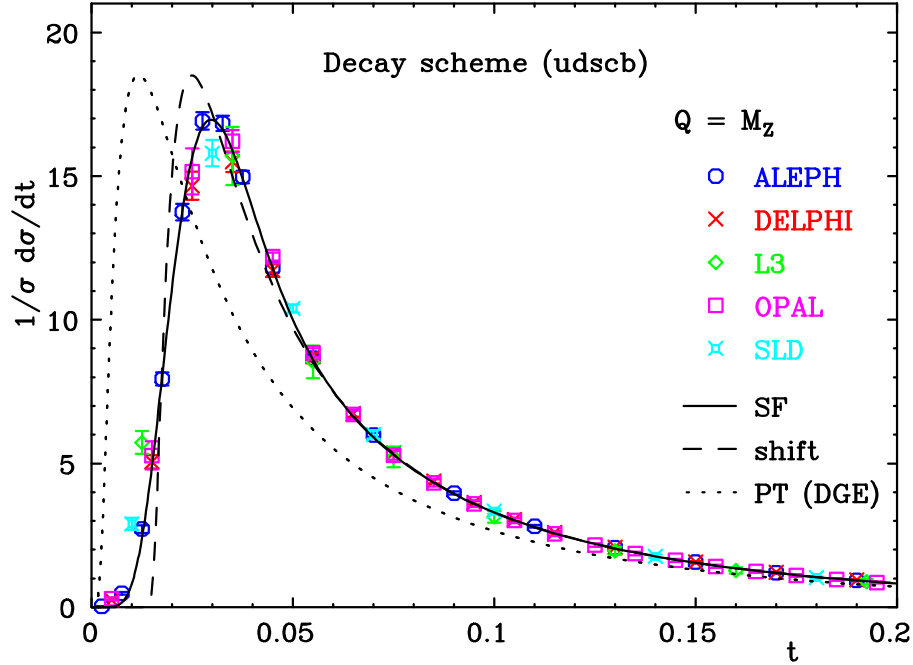}
\caption{Thrust data at ${M_{\rm Z}}$ in the decay scheme together
with the best shape-function-based fit (solid line) and a shift-based fit using the range $0.06\,{M_{\rm Z}}/Q < t< 0.30$ (dashed
line). The DGE perturbative calculation is also shown (dotted
line).} \label{fig:thrust_kcorr_mz}
\end{figure}

Figure~\ref{fig:thrust_kcorr_mz} shows the data at ${M_{\rm Z}}$ in
the decay scheme together with the best shape-function-based
fit~(\ref{eq:thrust_shaperes}) and a shift-based fit. For the
latter we use the range $0.06\,{M_{\rm Z}}/Q < t< 0.30$, which gives
the same $\alpha_s$ as the shape-function fit (see
Table~\ref{tab:tmin}). The figure illustrates clearly the
hierarchy between power corrections as a function of $t$. For
large $t$ the shape function is well approximated by a shift. In
the peak region the non-perturbative corrections with higher power
become important, but the effects are still small. Finally, below 
the peak all power corrections become equally important.

\begin{figure}[t]
\center \epsfig{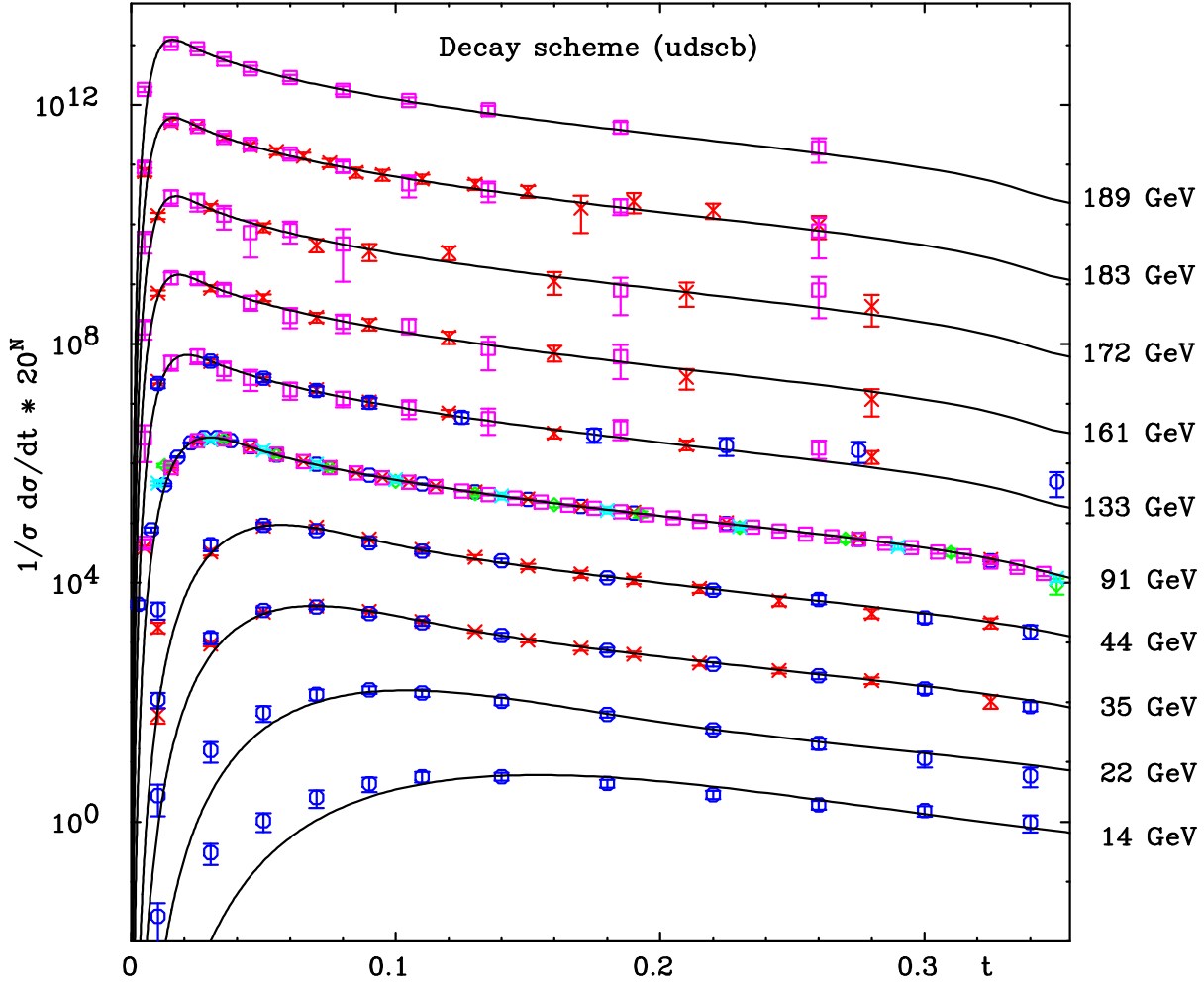}
\caption{Comparison of the thrust data in the decay scheme  with
the best fit shape function. Both the fit and the data have been
multiplied by a factor $20^N$, with $N=0\ldots 9$ for
$Q=14\ldots 189$ GeV.  Data points with errors larger than 100\%
are not shown.} \label{fig:thrust_kcorr_all}
\end{figure}

For illustration, the best shape-function-based
fit~(\ref{eq:thrust_shaperes}) is also shown at all energies, together with the
data, in Fig.~\ref{fig:thrust_kcorr_all}.
The contribution to the $\chi^2$ from the individual experiments
for the best shape-function-based fit~(\ref{eq:thrust_shaperes})
and the shift-based fit is given in Table~\ref{chi_sq_exp}. The
fits are good at all energies, except at $14$ GeV. This exception
is not surprising, since the effects of heavy quarks were not
taken into account.

To investigate the effects of heavy primary quarks we have also
performed the shape-function-based fit to the data which was
transformed according to~(\ref{eq:uds_transf}), so as to remove
heavy-quark events from the data sample. As before, this
transformation has been made using the {\sc Pythia} Monte Carlo model.
The results using a shape function with exponential fall-off, are
\begin{eqnarray}
{\rm D \; (uds): } &&
\begin{tabular}{ll}
$\alpha_{\overline{\mbox{\tiny MS}}}({M_{\rm Z}}) = 0.1092 \pm 0.0004$&  \\
$\lambda_1\bar{\Lambda} = 0.568\pm 0.013 \; \; {\rm GeV}$ &
$\lambda_2\bar{\Lambda}^2 = 0.005\pm 0.023 \; \; {\rm GeV}^2$ \\
$b_1 = 0.64 \pm  0.10$  &$b_2 \equiv 0$ \\
$q = 1.51  \pm 0.12 $&
\end{tabular}
\end{eqnarray}
with $\chi^2/{\rm dof}=0.63 $ for 225  points. This shows 
(cf. Table~\ref{tab:scheme}) that the effect of heavy quarks is
slightly larger in the peak region with respect to higher $t$, but it
is still small.

Considering the contribution to the $\chi^2$ from the individual
experiments, that of $14$~GeV becomes only slightly smaller
($14.1 \to 12.2$) whereas all the others increase. The Monte Carlo results
suggest that the effect of heavy quarks 
is small, except at 14 GeV, in agreement with~\cite{MovillaFernandez:2001ed}.
However, the change in $\chi^2$ raises doubts concerning the reliability of 
using the 
Monte Carlo together with Eq.~(\ref{eq:uds_transf}) to quantify this effect. 
It is clear that there is place for improving
the analysis in this respect.

Finally, we give the results of a fit with a shape function with
exponential fall-off to the data in the P scheme:
\begin{eqnarray}
\label{eq:thrust_shaperes_pscheme}
{\rm P \; (udscb): } &&
\begin{tabular}{ll}
 $\alpha_{\overline{\mbox{\tiny MS}}}({M_{\rm Z}}) = 0.1068 \pm 0.0004$&  \\
 $\lambda_1\bar{\Lambda} = 0.388\pm 0.013 \; \; {\rm GeV}$ &
$\lambda_2\bar{\Lambda}^2 = 0.069\pm 0.022 \; \; {\rm GeV}^2$ \\
 $b_1 = 0.47 \pm  0.05$  &$b_2 \equiv 0$ \\
 $q = 0.73  \pm 0.08 $&
\end{tabular}
\end{eqnarray}
with $\chi^2/{\rm dof}=0.67 $ for 225  points. This result agrees reasonably
well with the shift-based fit in Table~\ref{tab:scheme}. The fact that here, in
contrast to (\ref{eq:thrust_shaperes}), the non-perturbative parameter
$\lambda_2$ is significantly different from zero, $\lambda_2 \sim \lambda_1^2$,
is interesting. The differences between the 
non-perturbative parameters\footnote{It is expected that the
finite hadron mass affects also the higher power corrections. Note
that the modification of the distribution when changing the HMS is
more complicated than just a shift -- see
Fig.~\ref{fig:thrust_scheme_mz}.} $\lambda_n$ in
different HMS are associated with mass effects. It is rather clear that these
effects, at least their non-universal parts~\cite{Salam:2001bd}, do not have 
a ``perturbative'' origin, and therefore it is not surprising that they do 
not admit the renormalon power correction pattern.   

To summarize the fits with a shape function, the best fit in the decay scheme
is~(\ref{eq:thrust_shaperes}), which agrees well with the shift-based 
fit~(\ref{eq:thrust_shiftres}). The second central moment $\lambda_2$  in this
scheme is consistent  with zero, in accordance with the large $\beta_0$
renormalon pattern. This is not the case in other HMS. Similarly to the
shift-based fits, the only non-negligible  source of uncertainty in $\alpha_s$
is the choice of HMS.

\begin{table}[H]
\caption{Summary of the experimental data used and the
contribution to $\chi^2$ from the individual experiments (in the
decay scheme with all primary quarks). For the thrust, the data
were fitted using a shift ($0.06 \, {M_{\rm Z}}/Q < t < 0.30$) and a
shape function (SF) with exponential fall-off ($0.01 \,{M_{\rm Z}}/Q
< t < 0.30$), whereas for the heavy-jet mass ($0.01 \,{M_{\rm Z}}/Q
< \rho_H < 0.10$) the $\chi^2$ values were obtained with
$\alpha_{\overline{\mbox{\tiny MS}}}({M_{\rm Z}})=0.109$ and with the
shape function fixed according to the thrust analysis.
\label{chi_sq_exp}}
\begin{center}
\begin{tabular}{|lcc|cc|cc|cc|}
\hline
               &                     &        &  \multicolumn{2}{c|}{thrust \,\, shift} & \multicolumn{2}{c|}{thrust\,\,\,\,\, SF}  & \multicolumn{2}{c|}{heavy jet\,\,\, SF} \\
               &                     &        &  \multicolumn{2}{c|}{ } & \multicolumn{2}{c|}{ }  & \multicolumn{2}{c|}{($\alpha_s$ \& SF fixed)} \\
 Experiment    & Reference           & Q [GeV]&  $\chi^2$ &Points&$\chi^2$&Points&$\chi^2$& Points\\
\hline
  TASSO & \cite{TASSO,TASSOMASS}      &  14  &  0.00    &   0  & 14.12  &  6 &  1.59  &  1 \\
  TASSO & \cite{TASSO,TASSOMASS}      &  22  &  0.00    &   0  &  3.80  &  7 &  0.52  &  2 \\
   JADE & \cite{JADE}                 &  35  &  2.16    &   5  &  8.59  & 11 &  8.60  &  3 \\
  TASSO & \cite{TASSO,TASSOMASS}      &  35  &  1.44    &   3  &  3.94  &  8 &  0.61  &  3 \\
   JADE & \cite{JADE}                 &  44  &  4.24    &   6  &  4.82  & 11 &  4.82  &  3 \\
  TASSO & \cite{TASSO,TASSOMASS}      &  44  &  0.84    &   3  &  4.22  &  8 &  0.08  &  3 \\
  ALEPH & \cite{ALEPH91}              & 91.2 &  3.35    &   9  &  7.82  & 17 & 12.10  & 10 \\
 DELPHI & \cite{DELPHI91}             & 91.2 &  9.60    &  12  & 14.64  & 17 & 10.28  &  8 \\
     L3 & \cite{L391}                 & 91.2 &  3.67    &   7  &  3.90  &  9 &  6.15  &  5 \\
   OPAL & \cite{OPAL91}               & 91.2 & 10.90    &  24  & 20.42  & 29 & 66.23  & 10 \\
    SLD & \cite{SLD91}                & 91.2 &  3.48    &   5  &  5.19  &  7 & 11.69  &  2 \\
  ALEPH & \cite{ALEPH133}             & 133  &  1.58    &   6  &  1.60  &  8 &        &     \\
 DELPHI & \cite{DELPHI133_161_172_183}& 133  &  1.47    &   5  &  1.87  &  7 &  0.62  &  3 \\
   OPAL & \cite{OPAL133}              & 133  &  2.16    &   6  &  2.51  & 10 &  2.09  &  6 \\
 DELPHI & \cite{DELPHI133_161_172_183}& 161  &  3.47    &   6  &  4.81  &  7 &  1.06  &  3 \\
   OPAL & \cite{OPAL161}              & 161  &  1.77    &   7  &  1.95  & 10 &  1.70  &  6 \\
 DELPHI & \cite{DELPHI133_161_172_183}& 172  &  3.11    &   6  &  3.18  &  7 &  0.18  &  3 \\
   OPAL & \cite{OPAL172_183_189}      & 172  &  2.14    &   7  &  2.31  & 10 &  2.74  &  7 \\
 DELPHI & \cite{DELPHI133_161_172_183}& 183  &  4.22    &  14  &  4.84  & 16 &  0.62  &  7 \\
   OPAL & \cite{OPAL172_183_189}      & 183  &  0.58    &   8  &  0.62  & 10 &  2.04  &  7 \\
   OPAL & \cite{OPAL172_183_189}      & 189  &  0.94    &   8  &  1.40  & 10 &  2.32  &  7 \\
\hline
  Sum  &                              &      &  61.1    &  147 &  116.5 & 225& 136.0  & 99\\
\hline
\end{tabular}
\end{center}
\end{table}

\subsubsection{Fits with a modified logarithm}

To estimate the effects of the non-logarithmic terms, which are not included in the
resummation, we repeat the calculation using a modified logarithm,
$L=\ln(1/t-1)$~\cite{Catani:1991kz}. The results of fitting $\alpha_s$ and a shift
$\lambda_1$ using the modified logarithm in different HMS is summarized in
Table~\ref{tab:logmod}. Comparing with Table~\ref{tab:scheme} we see that the central
values of~$\alpha_s$ are~$\sim 2\%$ larger and $\lambda_1\bar{\Lambda}$ is $\sim 0.1$
GeV smaller, whereas the HMS dependence is unchanged. We also note that the $\chi^2$
values are \emph{significantly worse} when using a modified logarithm.

\begin{table}[t]
\caption{ Fits of $\alpha_s$ and $\lambda_1$ to the thrust data with a lower
cut $t_{\min}=0.05\,{M_{\rm Z}}/Q$ and an upper cut $t_{\max}=0.30$
in different hadron mass schemes using a modified logarithm,
$L=\ln(1/t-1)$.}
\begin{center}
\begin{tabular}{lcccc}
\tableline
 HMS & $\alpha_{\overline{\mbox{\tiny MS}}}({M_{\rm Z}})$ & $\lambda_1\bar{\Lambda}$ [GeV]& $\chi^2/{\rm dof} $ &  Points  \\
 M/P (udscb)  &  0.1096 $\pm$ 0.0005    &  0.25  $\pm$  0.03  & 0.70  & 159 \\
 E (udscb)    &  0.1104 $\pm$ 0.0005    &  0.32  $\pm$  0.02  & 0.62  & 159 \\
 D (udscb)    &  0.1107 $\pm$ 0.0005    &  0.55  $\pm$  0.02  & 0.50  & 159 \\
\tableline
\end{tabular}
\end{center}
\label{tab:logmod}
\end{table}

In addition to the HMS dependence we have also investigated the
dependence on the lower and upper cuts for the fit in the decay
scheme. Recall that with the unmodified logarithm we found no 
dependence on these cuts outside the peak region. 
However, with the modified logarithm we find a large $t_{\min}$  
dependence while keeping the
upper cut fixed at $t_{\max}=0.30$. As shown in
Table~\ref{tab:modmin}, changing the lower cut
from $0.05\,{M_{\rm Z}}/Q$ to $0.10\,{M_{\rm Z}}/Q$, the result for
$\alpha_s$ increases by~$4\%$  and $\lambda_1\bar{\Lambda}$
decreases by $\sim 0.2$ GeV. The values only stabilize for
$t_{\min}=0.07\,{M_{\rm Z}}/Q$, giving $\alpha_{\overline{\mbox{\tiny
MS}}}({M_{\rm Z}})=0.113\pm0.001$ and $\lambda_1\bar{\Lambda}=0.42
\pm 0.04$. Likewise we find that keeping the lower cut fixed at
$t_{\min}=0.07\,{M_{\rm Z}}/Q$ and varying the upper cut from
$0.15$ to $0.35$ the result for $\alpha_s$ changes by~$1.5\%$
and $\lambda_1\bar{\Lambda}$ by $\sim 0.05$ GeV.
\begin{table}[t]
\caption{Modified logarithm ($L=\ln(1/t-1)$) fits of 
$\alpha_s$ and $\lambda_1$ to the thrust data in the decay
scheme (udscb) as a function of the lower cut
$t_{\min}=t_{\min}({M_{\rm Z}}){M_{\rm Z}}/Q$ with the upper cut
fixed as $t_{\max}=0.30$.}
\begin{center}
\begin{tabular}{ccccc}
\tableline
$t_{\min}({M_{\rm Z}})$  & $\alpha_{\overline{\mbox{\tiny MS}}}({M_{\rm Z}})$ & $\lambda_1\bar{\Lambda}$ [GeV]& $\chi^2/{\rm dof} $ &  Points  \\
 0.03       &  0.1086 $\pm$ 0.0003    &  0.65  $\pm$  0.01  & 0.73  & 190 \\
 0.04       &  0.1090 $\pm$ 0.0003    &  0.64  $\pm$  0.01  & 0.72  & 173 \\
 0.05       &  0.1107 $\pm$ 0.0005    &  0.55  $\pm$  0.02  & 0.50  & 159 \\
 0.06       &  0.1119 $\pm$ 0.0006    &  0.48  $\pm$  0.03  & 0.44  & 147 \\
 0.07       &  0.1129 $\pm$ 0.0008    &  0.42  $\pm$  0.04  & 0.45  & 135 \\
 0.08       &  0.1134 $\pm$ 0.0009    &  0.39  $\pm$  0.05  & 0.44  & 124 \\
 0.09       &  0.1138 $\pm$ 0.0012    &  0.36  $\pm$  0.08  & 0.46  & 111 \\
 0.10       &  0.1146 $\pm$ 0.0016    &  0.31  $\pm$  0.10  & 0.48  & 104 \\
\tableline
\end{tabular}
\end{center}
\label{tab:modmin}
\end{table}

Even though the fits with a modified logarithm are significantly
worse than the fits with an unmodified logarithm these results can
be used to estimate the significance of the non-logarithmic terms.
Based on this we conclude that the effects of non-logarithmic
terms can be as large as $4\%$ in $\alpha_s$ (from
$\alpha_{\overline{\mbox{\tiny MS}}}({M_{\rm Z}})=0.109$ to
$0.113$). 
This is by far the dominant contribution to theoretical
uncertainty. We emphasize that this is a rough estimate and 
full NNLO calculations are essential to reduce the uncertainty. 

\subsection{Heavy jet mass distribution}

As explained in section~\ref{sec:applicability}, the approximation we use for
the  phase space breaks down completely at $\rho_H=1/6$. Contrary
to the case of the thrust, non-logarithmic corrections reflecting additional
kinematic constraints  may be important at rather small values of $\rho_H$. 
This will be reflected in the independent phenomenological analysis of the
heavy-jet mass distribution in section~3.2.1. We prefer to rely on the thrust
analysis as much as possible and compare between the non-perturbative 
corrections to the two observables only in the limited region where the calculation of the
heavy-jet mass distribution can be trusted.  This is the purpose of
section~3.2.2.

\subsubsection{Independent fits}

The purpose of this section is to confront the DGE resummed cross section and
the suggested parameterization of power corrections with the heavy-jet mass
distribution data and find how severe the limitations described in
section~\ref{sec:applicability} are in practice. 

We begin by fitting $\alpha_s$ and a shift of the perturbative 
distribution~(\ref{eq:shift}). This should be applicable far enough above  the
peak region. However, since the region where the calculation can be trusted is
strongly restricted from above, we end up with a rather narrow range. The upper
cut will be chosen as $\rho_{\max}=0.15$. We will show that the lower 
cut can be as small as $\rho_{\min}=0.03\,{M_{\rm Z}}/Q$.

The dependence on the HMS and the effect of heavy primary quarks are summarized
in Table~\ref{tab:hscheme}. First of all we note that the extracted values of
$\alpha_s$ are much smaller than for the thrust case. The reason is that our
approximation tends  to overestimate the higher order perturbative coefficients
at large $\rho_H$: an example is provided by the NLO coefficient shown in
Fig.~\ref{fig:nlo}. This statement will be further supported by studying the
dependence on the fitting range~(Table~\ref{tab:hmax}). It should also be kept
in mind that, since the distribution is normalized,
overestimating the cross section  at large $\rho_H$ is compensated by 
underestimating it at low $\rho_H$. For the NLO coefficient the turnover  
point is $\rho_H \sim 0.05$. 
\begin{table}[t]
\caption{ Fits of $\alpha_s$ and $\lambda_1$ to the heavy-jet mass data with a lower
cut of $\rho_{\min}=0.03\,{M_{\rm Z}}/Q$ and an upper cut of
$\rho_{\max}=0.15$ in different HMS.}
\begin{center}
\begin{tabular}{lcccc}
\tableline
 HMS & $\alpha_{\overline{\mbox{\tiny MS}}}({M_{\rm Z}})$ & $\lambda_1\bar{\Lambda}$ [GeV]& $\chi^2/{\rm dof} $ &  Points  \\
 P (udscb)  &  0.1009 $\pm$ 0.0006    &  0.58  $\pm$  0.03  & 0.42  & 87 \\
 P (uds)    &  0.1012 $\pm$ 0.0008    &  0.55  $\pm$  0.04  & 0.41  & 87 \\
 E (udscb)  &  0.1014 $\pm$ 0.0007    &  0.69  $\pm$  0.04  & 0.44  & 87 \\
 E (uds)    &  0.1018 $\pm$ 0.0007    &  0.65  $\pm$  0.04  & 0.43  & 87 \\
 D (udscb)  &  0.1019 $\pm$ 0.0007    &  0.88  $\pm$  0.03  & 0.50  & 87 \\
 D (uds)    &  0.1020 $\pm$ 0.0007    &  0.85  $\pm$  0.03  & 0.50  & 87 \\
 M (udscb)  &  0.1026 $\pm$ 0.0006    &  1.03  $\pm$  0.03  & 0.56  & 87 \\
\tableline
\end{tabular}
\end{center}
\label{tab:hscheme}
\end{table}

Since the values of $\alpha_s$ obtained here are so different from the thrust
case, it is no surprise that the values of the shift $\lambda_1$ differ by
much as well. In any case, a meaningful comparison  of these parameters requires
the same value of $\alpha_s$ to be used. We will come back to such a comparison 
below.
 
\begin{figure}[t]
\center \epsfig{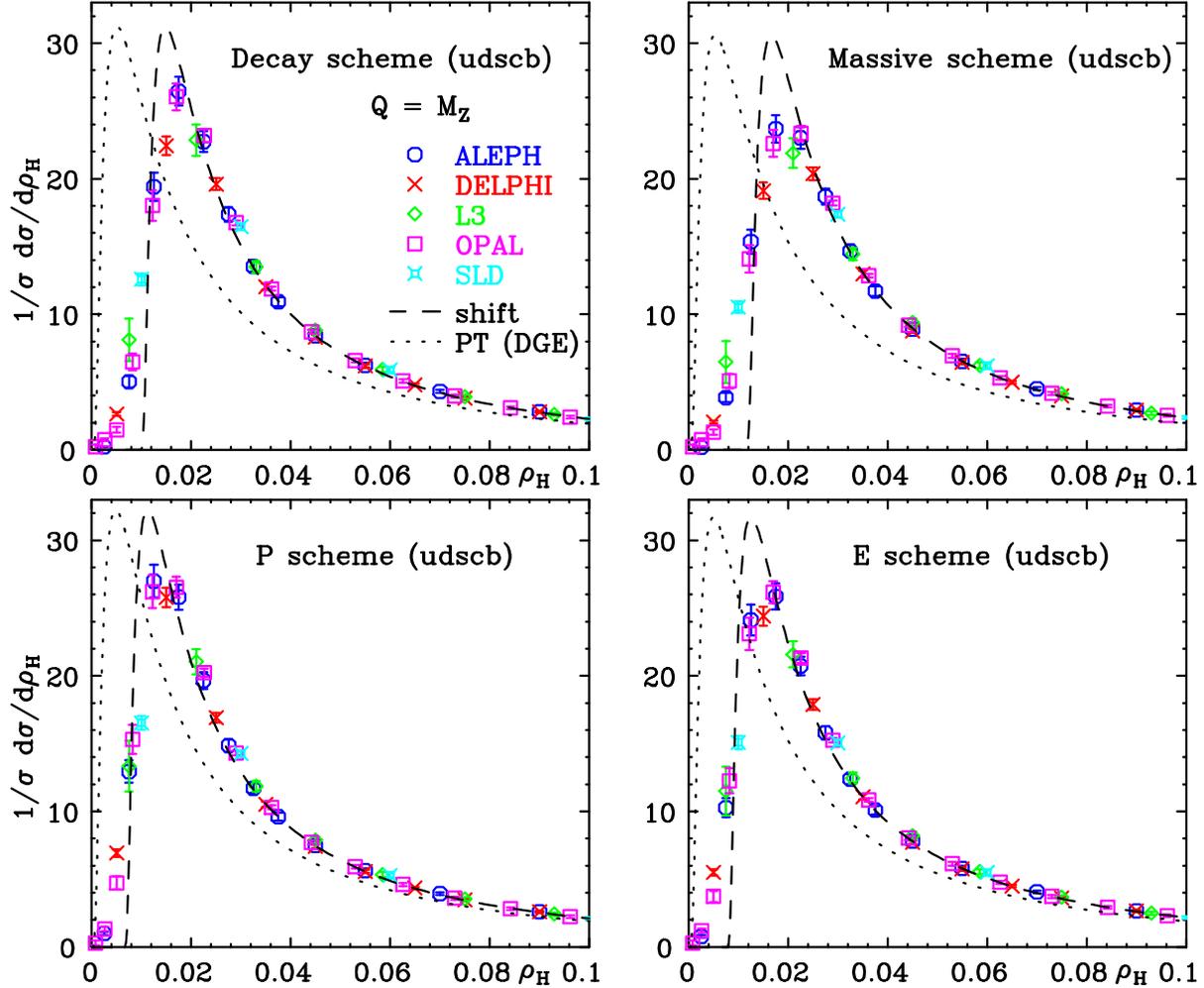}
\caption{Heavy jet mass data at ${M_{\rm Z}}$ in the peak region
 together with shift-based fits using the range
$0.03\,{M_{\rm Z}}/Q < \rho_H< 0.15$ (dashed line) in different
HMS. The perturbative calculation is also shown
(dotted line).} \label{fig:rhoh_scheme_mz}
\end{figure}

We note that, similarly to the thrust case, the HMS dependence and the effects
of primary heavy quarks on the value of $\alpha_s$ is small. The largest
difference is  between the P and the massive schemes. The differences between
the two can be clearly seen in Fig.~\ref{fig:rhoh_scheme_mz}, where the fit
results at ${M_{\rm Z}}$  in the different HMS are shown together with the
data\footnote{The figure also illustrates the agreement between the different
data sets in the peak region (in a given HMS) once they are all corrected to be
for all particles.}. In the massive scheme the contribution from the hadron
masses themselves shifts the distribution to the right. In general, as in the
thrust case, the HMS transformations modify both the position and the height of
the peak. Note, however, that the decay and massive schemes are quite similar
for the heavy-jet mass. The reason is that the only difference between the two
is the definition of the thrust axis. 

Next, we investigate the variation of the result as a function of the fitting
range in the decay scheme. From Tables~\ref{tab:hmin} and~\ref{tab:hmax}, it
can be concluded, based on the same criteria applied in the thrust analysis, that
the maximal range that can be used is~$0.03\,{M_{\rm Z}}/Q \leq \rho_h \leq
0.15$. However, contrary to the thrust case, here we identify a clear trend: 
as the fitting range shifts towards larger values of $\rho_H$, the value of
$\alpha_s$ decreases. Thus, there is a systematic discrepancy between the shape
of the theoretical distribution and the data. From Table~\ref{tab:hmax} it is
also clear that the fit deteriorates quickly for $\rho_{\max} > 1/6$. This is a
direct consequence of the missing kinematic constraints in the calculation~(see
Fig.~\ref{fig:nlo}).
\begin{table}[t]
\caption{ Fits of $\alpha_s$ and $\lambda_1$ to the heavy-jet mass data in the decay scheme (udscb) as a function of the lower cut
$\rho_{\min}=\rho_{\min}({M_{\rm Z}}){M_{\rm Z}}/Q$, with the upper
cut fixed as $\rho_{\max}=0.15$.}
\begin{center}
\begin{tabular}{ccccc}
\tableline
$\rho_{\min}({M_{\rm Z}})$  & $\alpha_{\overline{\mbox{\tiny MS}}}({M_{\rm Z}})$ & $\lambda_1\bar{\Lambda}$ [GeV]& $\chi^2/{\rm dof} $ &  Points  \\
 0.02       &  0.1009 $\pm$ 0.0005    &  0.94  $\pm$  0.02  & 0.62  &  106 \\
 0.03       &  0.1019 $\pm$ 0.0007    &  0.88  $\pm$  0.03  & 0.50  &  87 \\
 0.04       &  0.1023 $\pm$ 0.0010    &  0.86  $\pm$  0.05  & 0.44  &  72 \\
 0.05       &  0.1012 $\pm$ 0.0014    &  0.95  $\pm$  0.09  & 0.38  &  60 \\
 0.06       &  0.1000 $\pm$ 0.0019    &  1.05  $\pm$  0.14  & 0.34  &  48 \\
 0.07       &  0.0990 $\pm$ 0.0031    &  1.12  $\pm$  0.24  & 0.36  &  43 \\
\tableline
\end{tabular}
\end{center}
\label{tab:hmin}
\end{table}
\begin{table}[t]
\caption{ Fits of $\alpha_s$ and $\lambda_1$ to the heavy-jet mass 
data in the decay
scheme (udscb) as a function of the upper cut $\rho_{\max}$, with
the lower cut fixed as $\rho_{\min}=0.03\,{M_{\rm Z}}/Q$.}
\begin{center}
\begin{tabular}{ccccc}
\tableline
$\rho_{\max}$  & $\alpha_{\overline{\mbox{\tiny MS}}}({M_{\rm Z}})$ & $\lambda_1\bar{\Lambda}$ [GeV]& $\chi^2/{\rm dof} $ &  Points  \\
 0.10       &  0.1028 $\pm$ 0.0009    &  0.86  $\pm$  0.04  & 0.31  &  58 \\
 0.11       &  0.1027 $\pm$ 0.0009    &  0.87  $\pm$  0.04  & 0.30  &  60 \\
 0.12       &  0.1024 $\pm$ 0.0008    &  0.87  $\pm$  0.03  & 0.38  &  69 \\
 0.13       &  0.1023 $\pm$ 0.0008    &  0.87  $\pm$  0.03  & 0.40  &  76 \\
 0.14       &  0.1021 $\pm$ 0.0007    &  0.88  $\pm$  0.03  & 0.43  &  84 \\
 0.15       &  0.1019 $\pm$ 0.0007    &  0.88  $\pm$  0.03  & 0.50  &  87 \\
 0.16       &  0.1014 $\pm$ 0.0006    &  0.90  $\pm$  0.03  & 0.57  &  95 \\
 0.18       &  0.1006 $\pm$ 0.0006    &  0.92  $\pm$  0.03  & 0.93  &  103 \\
 0.20       &  0.0995 $\pm$ 0.0005    &  0.95  $\pm$  0.03  & 1.39  &  114 \\
\tableline
\end{tabular}
\end{center}
\label{tab:hmax}
\end{table}

To demonstrate the importance of using data that have been corrected to be for
all particles rather than for charged ones only, we made a separate fit 
to ALEPH and
DELPHI data at ${M_{\rm Z}}$ in the massive scheme ($\rho_{\min}=0.03\,
M_{\rm Z}/Q$ and $\rho_{\max}=0.15$). The result changes from
$\alpha_{\overline{\mbox{\tiny MS}}}({M_{\rm Z}})  =
0.1039  \pm 0.0009$ and $\lambda_1\bar{\Lambda}= 0.95  \pm  0.04$ GeV with
$\chi^2/{\rm dof} =0.22$ for all particles, to 
$\alpha_{\overline{\mbox{\tiny MS}}}({M_{\rm Z}})  = 0.1086 \pm 0.0007$ and
$\lambda_1\bar{\Lambda}= 0.88  \pm  0.03$ GeV with $\chi^2/{\rm dof} =0.99$ for
charged particles. Thus, contrary to the case of the
thrust distribution, the difference in $\alpha_s$ is very significant. The
agreement between the values of $\alpha_s$ between the thrust and heavy-jet
mass distributions when fitting charged particles only should be regarded as
accidental for the reason already given. Note that $\chi^2$ is much better for all particles than for charged particles only.

We have also performed fits to the experimental data using a 
shape function with exponential fall-off. For the lower cut
$\rho_{\min}=0.01\,{M_{\rm Z}}/Q$ and the upper cut
$\rho_{\max}=0.15$, we find
\begin{eqnarray}
{\rm D \; (udscb): } &&
\begin{tabular}{ll}
$\alpha_{\overline{\mbox{\tiny MS}}}({M_{\rm Z}}) = 0.1020 \pm 0.0005$&  \\
$\lambda_1\bar{\Lambda} = 0.84\pm 0.03 \; \; {\rm GeV}$ &
$\lambda_2\bar{\Lambda}^2 = -0.085\pm 0.032 \; \; {\rm GeV}^2$ \\
$b_1 = 0.70 \pm  0.10$  &$b_2 \equiv 0$ \\
$q = 0.50  \pm 0.77 $&
\end{tabular}
\end{eqnarray}
with $\chi^2/{\rm dof}=0.62 $ for 132  points. For comparison we repeated the 
fit in the P scheme:
\begin{eqnarray}
{\rm P \; (udscb): } &&
\begin{tabular}{ll}
$\alpha_{\overline{\mbox{\tiny MS}}}({M_{\rm Z}}) = 0.1008 \pm 0.0006$&  \\
$\lambda_1\bar{\Lambda} = 0.551 \pm 0.012  \; \; {\rm GeV}$ &
$\lambda_2\bar{\Lambda}^2 = -0.089   \pm 0.015  \; \; {\rm GeV}^2$ \\
$b_1 = 0.37  \pm  0.02 $  &$b_2 \equiv 0$ \\
$q =  10^{-7}   \pm 0.11   $&
\end{tabular}
\end{eqnarray}
with $\chi^2/{\rm dof}=0.47 $ for 132  points. The results of
these fits are in good agreement with the shift-based fits. This
gives further support to the use of the lower cut
$\rho_{\min}=0.03\,{M_{\rm Z}}/Q$ when fitting with a shift.

\begin{figure}[t]
\center
\epsfig{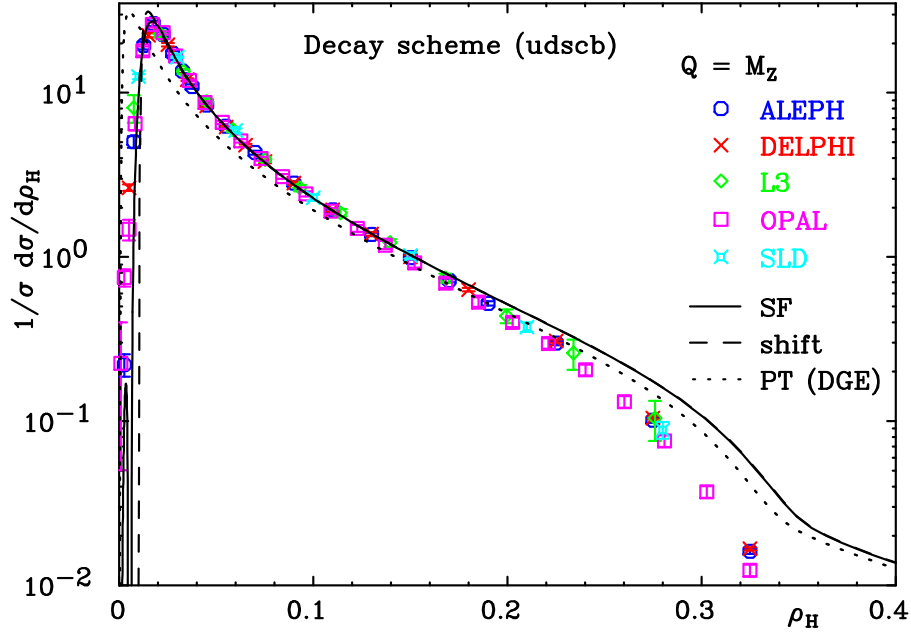}
\caption{Heavy jet mass data at ${M_{\rm Z}}$ in the decay scheme
for all primary quarks together with the results of the fits with
a shape function in the range $0.01\,{M_{\rm Z}}/Q < \rho_H <0.15$
(solid line) and a shift in the range $0.03\,{M_{\rm Z}}/Q < \rho_H
<0.15$ (dashed line). The perturbative calculation is also shown
(dotted line).} \label{fig:rhoh_kcorr_mz}
\end{figure}

The results of the best fits with a shape function and with a shift (with
lower cut $\rho_{\min}=0.03\,{M_{\rm Z}}/Q$) in the decay scheme are
shown together with the data at ${M_{\rm Z}}$ in Fig.~\ref{fig:rhoh_kcorr_mz}. 
The range shown is extended much above the upper cut 
in order to show the discrepancy discussed above 
between the calculation and the data.

Since the precision of the resummation formula deteriorates as $\rho_H$ is
increased, we have also tried to decrease the upper cut
$\rho_{\max}$ when fitting with a shape function. Limiting the fit
to include just the peak region by choosing $\rho_{\max}=0.06$ 
(the fit does not converge for lower~$\rho_{\max}$), the result is
\begin{eqnarray}
{\rm D \; (udscb): } &&
\begin{tabular}{ll}
$\alpha_{\overline{\mbox{\tiny MS}}}({M_{\rm Z}}) = 0.1038 \pm 0.0025$&  \\
$\lambda_1\bar{\Lambda} = 0.78\pm 0.09 \; \; {\rm GeV}$ &
$\lambda_2\bar{\Lambda}^2 = -0.057\pm 0.057 \; \; {\rm GeV}^2$ \\
$b_1 = 0.45 \pm  0.05$  &$b_2 \equiv 0$ \\
$q = 0.33  \pm 0.45 $&
\end{tabular}
\end{eqnarray}
with $\chi^2/{\rm dof}=0.55 $ for 62  points. This shows that an analysis of
the heavy-jet mass data restricted to the peak region is closer to the result
from the thrust analysis. In addition, one identifies a trend for the central
value of $\alpha_s$, which is similar to the shift-based fits: the lower the
upper cut is, the higher $\alpha_s$ gets.

Finally, we have fitted $\alpha_s$ and a shift $\lambda_1$ to the heavy-jet
mass data with a modified logarithm, $L=\ln(1/\rho_H-1)$. The effect of this
modification is very small, because of the limited range used in the fit.  As
discussed in section~(\ref{sec:applicability}), this modification does not at
all represent the magnitude of genuine non-logarithmic terms in this case.
 
To summarize the independent fits to the heavy-jet mass distribution, we find
smaller values of $\alpha_s$, and corresponding larger values of $\lambda_1$,
than those obtained in the thrust analysis. We understand this discrepancy by the
limited applicability of the phase-space approximation. We have
established that this primarily affects the distribution at larger values of
$\rho_H$. This is reflected for example in the fact that the best fit value of
$\alpha_s$ decreases as the fitting range shifts to the right.  Owing to the
limitations of the currently available calculations, it is not possible  to
extract $\alpha_s$ reliably from the heavy-jet mass data. A full NNLO
calculation can change the situation. 

\subsubsection{Comparison of the heavy-jet mass with thrust-based predictions}

While the available calculation of the heavy-jet mass is not accurate far
enough  above the peak region to provide an independent measurement of
$\alpha_s$, the main advantage of DGE is that it provides a solid basis for the
analysis of power corrections in the peak region~itself. 

The purpose of this section is to test our assumptions concerning power
corrections to the thrust and the heavy-jet mass distributions. This is done  
both by fitting the heavy-jet mass distribution with a fixed $\alpha_s$, which
is set according to the best fit of the thrust, and comparing the power 
corrections parameters $\lambda_n$, as well as by using the best fits of the
thrust distribution to determine all the parameters and confront the {\em
calculated} heavy-jet mass distribution with the data. 

We start by fitting a shift $\lambda_1$ of the perturbative heavy-jet mass
distribution, for a fixed value of $\alpha_s$ based on the thrust fits
(Table~\ref{tab:scheme}). The results are summarized in Table~\ref{tab:afix},
together with the best fit values of $\lambda_1$  obtained in the thrust case
for easy comparison.  The agreement between the two observables is quite
reasonable, except for the massive scheme, which is clearly disfavoured.  We
stress, however, that the heavy-jet mass results are quite  sensitive to the
fitting range and that the $\chi^2$ are much worse than in the fit where
$\alpha_s$ is free. This is especially true for the data at ${M_{\rm Z}}$ where
$\chi^2$/point $\gsim 2$.  The sensitivity to the upper and lower cuts is
summarized in Tables \ref{ts1} and \ref{ts2}, respectively. Examining Table
\ref{ts1} one can get the impression that the agreement of the shift parameters
between the two observables shown in Table \ref{tab:afix} is misleading,
since the values strongly depend on the lower cut. Nevertheless, based on the
shape-function fits presented below, the comparison in Table \ref{ts1} is
reliable: the shape-function-based fit values of $\lambda_1$ are consistent
with those obtained with a shift, provided that the lower cut for the latter 
is $\rho_{\min}=0.03\,{M_{\rm Z}}/Q$.
\begin{table}[t]
\caption{ Fits of $\lambda_1$ to the heavy-jet mass data with a lower cut
$\rho_{\min}=0.03\,{M_{\rm Z}}/Q$ and an upper cut $\rho_{\max}=0.15$
in different HMS using
$\alpha_{\overline{\mbox{\tiny MS}}}({M_{\rm Z}})$ from the thrust
analysis as input.}
\begin{center}
\begin{tabular}{lcccc|c}
\tableline
 HMS & $\alpha_{\overline{\mbox{\tiny MS}}}({M_{\rm Z}})$ & $\lambda_1\bar{\Lambda}$
 [GeV]& $\chi^2/{\rm dof} $ &  Points  & $(\lambda_1\bar{\Lambda})_{\rm thrust}$ [GeV]\\
 P (udscb)  &  0.1072     &  0.28  $\pm$  0.02  & 1.17  & 87 & 0.35 $\pm$  0.02\\
 E (udscb)  &  0.1080     &  0.38  $\pm$  0.02  & 1.32  & 87 & 0.42 $\pm$  0.02\\
 D (udscb)  &  0.1086     &  0.59  $\pm$  0.02  & 1.57  & 87 & 0.63 $\pm$  0.02\\
 M (udscb)  &  0.1072     &  0.85  $\pm$  0.01  & 1.15  & 87 & 0.35 $\pm$  0.02\\
\tableline
\end{tabular}
\end{center}
\label{tab:afix}
\end{table}
\begin{table}[t]
\caption{ 
Fits of a shift with a fixed $\alpha_{\overline{\mbox{\tiny MS}}}({M_{\rm Z}}) = 0.1086$ to the heavy-jet mass data in the decay scheme (udscb)
as a function of the
lower cut $\rho_{\min}=\rho_{\min}({M_{\rm Z}})\,{M_{\rm Z}}/Q$, with a fixed upper cut
$\rho_{\max}=0.15$.}
\begin{center}
\begin{tabular}{ccccc} 
\tableline
$\rho_{\min}({M_{\rm Z}})$  & $\lambda_1\bar{\Lambda}$ [GeV]&  $\chi^2/{\rm dof} $ &  Points  \\  
 0.02           &  0.70  $\pm$  0.01  & 3.16  &  106 \\
 0.03           &  0.59  $\pm$  0.02  & 1.57  &  87 \\
 0.04           &  0.51  $\pm$  0.02  & 0.98  &  72 \\
 0.05           &  0.45  $\pm$  0.03  & 0.79  &  60 \\
\tableline
\end{tabular}
\end{center}
\label{ts1}
\end{table}
\begin{table}[t]
\caption{ 
Fits of a shift with a fixed $\alpha_{\overline{\mbox{\tiny MS}}}({M_{\rm Z}}) = 0.1086$ to the heavy-jet mass data in the decay scheme (udscb)
as a function of the
upper cut $\rho_{\max}$ with a fixed lower cut $\rho_{\min}=0.03\,{M_{\rm Z}}/Q$.}
\begin{center}
\begin{tabular}{ccccc} 
\tableline
$\rho_{\max}$   & $\lambda_1\bar{\Lambda}$ [GeV]& $\chi^2/{\rm dof} $ &  Points  \\  
 0.10           &  0.64  $\pm$  0.02  &  0.95  &  58 \\
 0.11           &  0.63  $\pm$  0.02  &  1.00  &  60 \\
 0.12           &  0.61  $\pm$  0.02  &  1.27  &  69 \\
 0.13           &  0.61  $\pm$  0.02  &  1.26  &  76 \\
 0.14           &  0.60  $\pm$  0.02  &  1.47  &  84 \\
 0.15           &  0.59  $\pm$  0.02  &  1.57  &  87 \\
 0.16           &  0.58  $\pm$  0.02  &  1.94  &  95 \\
\tableline
\end{tabular}
\end{center}
\label{ts2}
\end{table}

\begin{figure}[t]
\center \epsfig{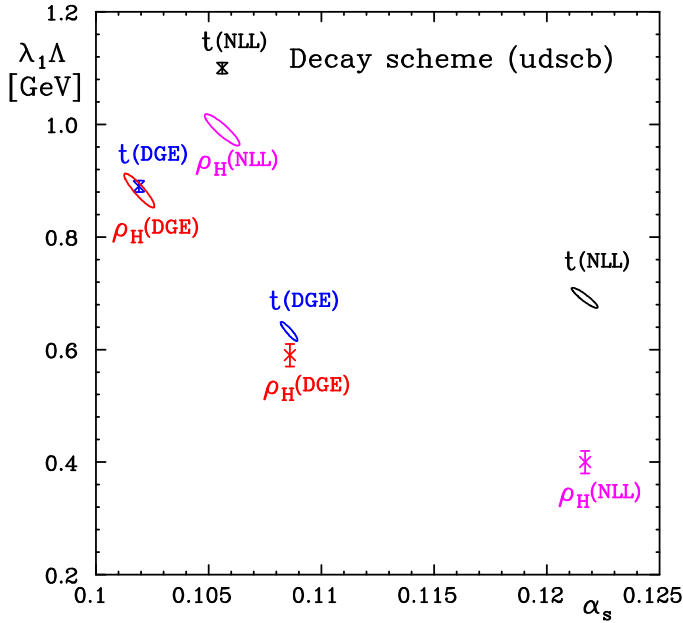}
\caption{The results of fitting $\alpha_s$ and a shift $\lambda_1$
to the thrust and heavy-jet mass data in the decay scheme using
DGE or NLL resummation (1-$\sigma$ contours). Also shown are the
results of fitting $\lambda_1$ for one of the observables fixing
$\alpha_s$ based on the other (crosses). The ranges used are
$0.05\,{M_{\rm Z}}/Q <t < 0.30$ and $0.03\,{M_{\rm Z}}/Q < \rho_H
<0.15$.} 
\label{fig:correlation}
\end{figure}

To compare between the thrust and heavy-jet mass fits and to illustrate the
correlation between $\alpha_s$ and $\lambda_1$, Fig.~\ref{fig:correlation}
shows the results of fitting the two parameters to the thrust and heavy-jet
mass data in the decay scheme as 1-$\sigma$ contours. The figure also shows
the results of fitting $\lambda_1$ for one of the observables when fixing
$\alpha_s$ based on the other. We notice that the heavy-jet mass and thrust
have very similar correlations between~$\alpha_s$ and~$\lambda_1$. In addition,
the figure shows the results of the same exercise using NLL resummation,
instead of DGE. Contrary to the DGE-based fits, in the NLL case there is no
agreement between the thrust and heavy-jet mass results when~$\alpha_s$ is fixed. This shows that important contributions are missed in the
NLL resummation. In addition, the magnitude of the non-perturbative 
corrections  for a given~$\alpha_s$ is, of course, smaller in the case of DGE.

A more stable result for the non-perturbative parameters  under variations of
the fitting range is obtained when including the peak region and  fitting with
a shape function. Fitting a shape function with an exponential fall-off in the
decay scheme, where $\alpha_s$ is fixed based on the corresponding thrust
analysis and the range is 
\hbox{$0.01\,{M_{\rm Z}}/Q < \rho_H < 0.10$}, yields 
\begin{eqnarray}
\label{D_rho_fixed_alpha}
{\rm D \; (udscb): } &&
\begin{tabular}{ll}
$\alpha_{\overline{\mbox{\tiny MS}}}({M_{\rm Z}}) \equiv 0.1090\; \; {\rm fixed}$&  \\
$\lambda_1\bar{\Lambda} = 0.600\pm 0.007 \; \; {\rm GeV} $ &
$\lambda_2\bar{\Lambda}^2 = 0.080\pm 0.019 \; \; {\rm GeV}^2$ \\
$b_1 = 0.50 \pm  0.01$  &$b_2 \equiv 0$ \\
$q = 2\cdot 10^{-8}  \pm 0.04 $&
\end{tabular}
\end{eqnarray}
with $\chi^2/{\rm dof}=0.93 $ for 99  points.  The reduced sensitivity of the
shape-function-based fit to the lower and upper  cuts is  shown in
Tables~\ref{tsf1} and~\ref{tsf2}, respectively. Note that fitting with a shape
function only makes sense if the peak region is included. Therefore, the most
relevant entries in Table~\ref{tsf1} are the first two lines.  The
agreement with the thrust result~(\ref{eq:thrust_shaperes}) for $\lambda_1$ is
very good. As for $\lambda_2$, 
although the $\rho_H$ fit results are not as stable with respect to the lower cut 
as those of the thrust (cf. Table~\ref{tab:tminsf}), it seems that the
$\rho_H$ fit in the decay scheme, as that of the thrust,
prefers a small~$\lambda_2$.  
For the parameter $b_1$ the agreement between the two results is also
good, but this is not the case for $q$. It should also be noted that the value
of $q$ in (\ref{D_rho_fixed_alpha}) is at the limit of the allowed range
imposed on the fit.
\begin{table}[t]
\caption{ 
Fits of a shape function with a fixed $\alpha_{\overline{\mbox{\tiny MS}}}({M_{\rm Z}}) = 0.1090$ to the heavy-jet mass data in the decay scheme (udscb) as a function of the
lower cut $\rho_{\min}=\rho_{\min}({M_{\rm Z}})\,{M_{\rm Z}}/Q$ with the upper cut
$\rho_{\max}=0.10$
fixed.}
\begin{center}
\begin{tabular}{ccccc} 
\tableline
$\rho_{\min}({M_{\rm Z}})$  & $\lambda_1\bar{\Lambda}$ [GeV]& $\lambda_2\bar{\Lambda}^2$ [GeV$^2$]& $\chi^2/{\rm dof} $ &  Points  \\  
 0.01           &  0.60   $\pm$  0.01  &  0.08    $\pm$ 0.02    & 0.93   &  99  \\
 0.02           &  0.60   $\pm$  0.02  & -0.01    $\pm$ 0.02    & 0.74   &  75  \\
 0.03           &  0.57   $\pm$  0.01  & -0.20    $\pm$ 0.07    & 0.37   &  58  \\
 0.04           &  0.39   $\pm$  0.20  & -0.29    $\pm$ 0.18    & 0.26   &  45  \\
\tableline
\end{tabular}
\end{center}
\label{tsf1}
\end{table}
\begin{table}[t]
\caption{ 
Fits of a shape function with a fixed $\alpha_{\overline{\mbox{\tiny MS}}}({M_{\rm Z}}) = 0.1090$ to the heavy-jet mass data in the decay scheme (udscb) as a function of the
upper cut $\rho_{\max}$ with the
lower cut $\rho_{\min}=0.01\,{M_{\rm Z}}/Q$ fixed.}
\begin{center}
\begin{tabular}{ccccc} 
\tableline
$\rho_{\max}$   & $\lambda_1\bar{\Lambda}$ [GeV]& $\lambda_2\bar{\Lambda}^2$ [GeV$^2$]& $\chi^2/{\rm dof} $ &  Points  \\  
 0.05           &  0.63  $\pm$  0.01  & 0.08  $\pm$ 0.02   & 0.58  &  52 \\
 0.10           &  0.60  $\pm$  0.01  & 0.08  $\pm$ 0.02   & 0.93  &  99 \\
 0.15           &  0.58  $\pm$  0.01  & 0.07  $\pm$ 0.02   & 1.45  &  132 \\
\tableline
\end{tabular}
\end{center}
\label{tsf2}
\end{table}

Finally, as an ultimate test of our assumptions, we fix both $\alpha_s$ 
and the shape function based on the thrust distribution best fit and  examine
the agreement of the calculated heavy-jet mass distribution with the  data.
This gives a $\chi^2/{\rm dof}=1.37 $ for $99$ points in the range $0.01\,{\rm
M_Z}/Q < \rho_H < 0.10$. The contribution to this $\chi^2$ from each experiment
is shown in Table~\ref{chi_sq_exp}. It is worth noting that for most of the
data sets the $\chi^2$ is good ($\chi^2 \lsim 1$). The only exceptions are JADE
at $35$ GeV, OPAL at ${M_{\rm Z}}$ and SLD. Excluding these data sets the
$\chi^2$ becomes $\chi^2/{\rm dof}=0.60$.

\begin{figure}[t]
\center \epsfig{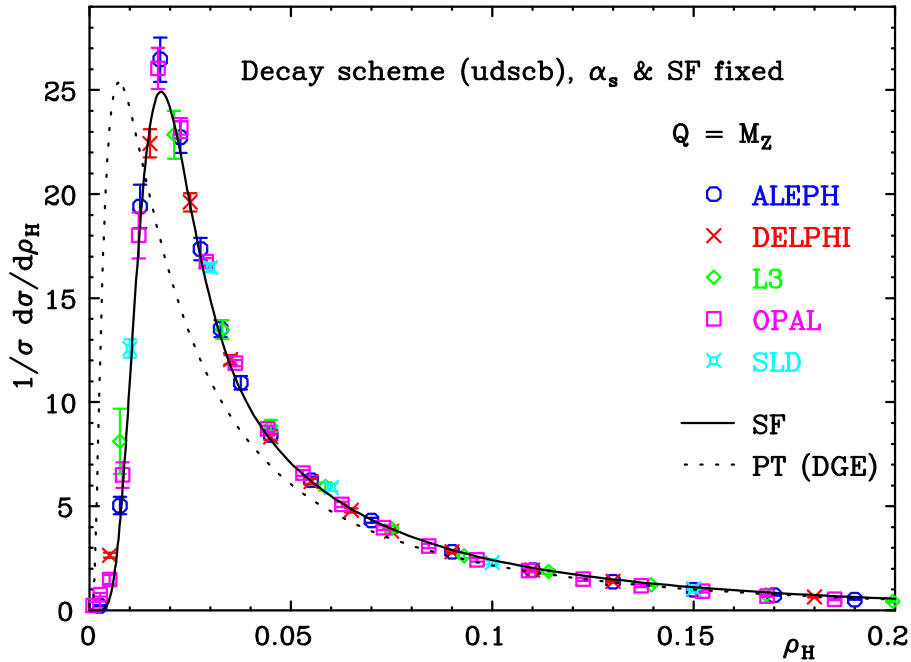}
\caption{Comparison of the heavy-jet mass data at ${M_{\rm Z}}$ in
the decay scheme with the theoretical prediction based on the
determination of $\alpha_s$ and the shape function from the thrust
analysis. } \label{fig:rhoh_kcorr_mz_allfix}
\end{figure}

\begin{figure}[t]
\center \epsfig{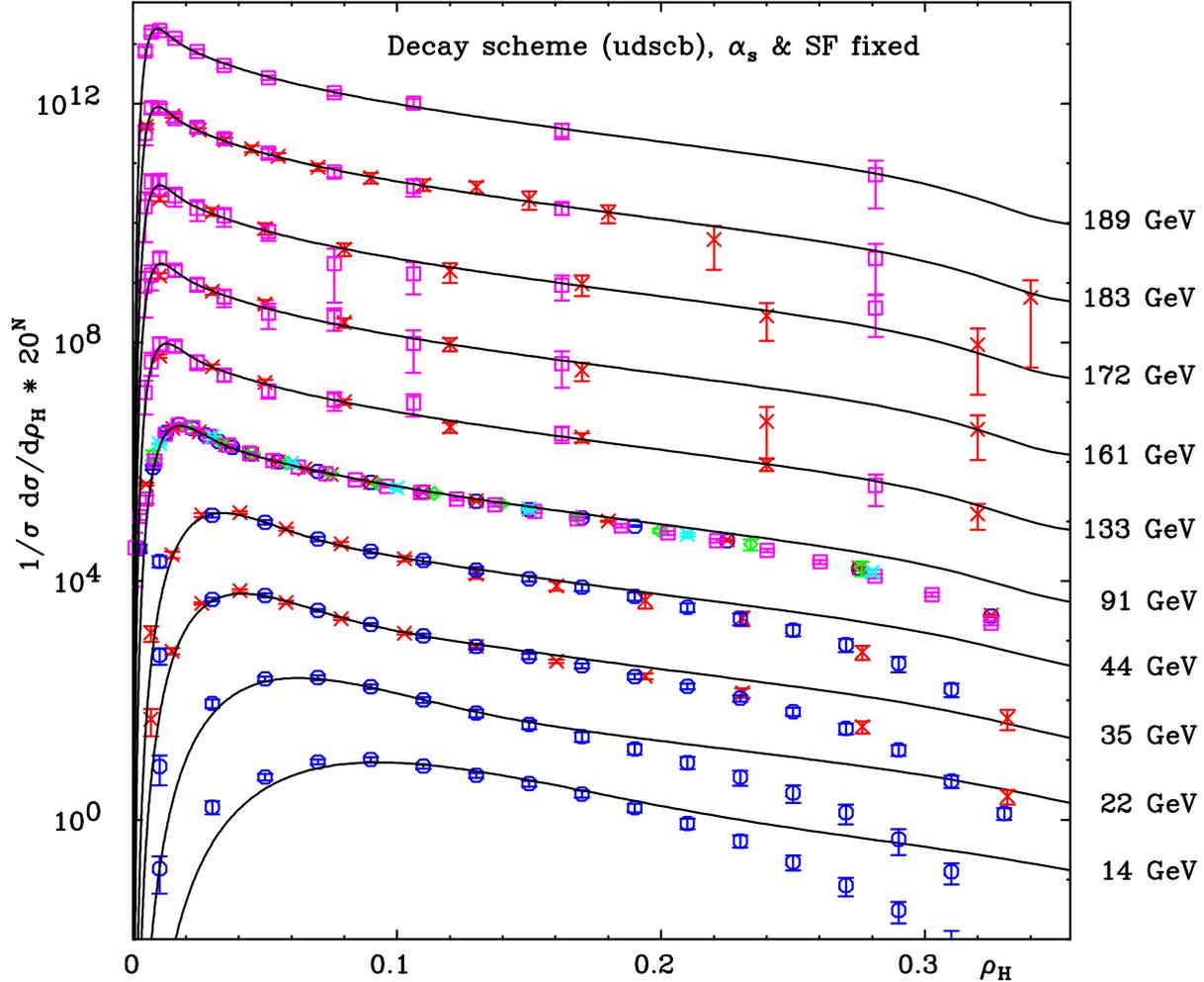}
\caption{Comparison of the heavy-jet mass data  in the decay
scheme with the theoretical prediction based on the determination
of $\alpha_s$ and the shape function from the thrust analysis.
Both the theoretical prediction and the data have been multiplied
by a factor $20^N$, with $N=0\ldots 9$ for $Q=14\ldots 189$ GeV.
Data points with errors larger than $100~\%$ are not shown.}
\label{fig:rhoh_all}
\end{figure}

Figure~\ref{fig:rhoh_kcorr_mz_allfix} shows the data at ${M_{\rm Z}}$
together with the prediction based on the thrust analysis. The
general agreement between the two is good. A closer look shows
that the theory curve systematically underestimates the data up to
$\rho_H\simeq 0.06$ and overestimates it to the right of this
point. 
In Fig.~\ref{fig:rhoh_all} we show the prediction based on the
thrust analysis together with the heavy-jet mass data at all
energies. Here we see that the discrepancy between the predicted distribution 
and the $\rho_H$ data increases as the energy is decreased. 
This is natural, since the larger value of
$\alpha_s$ makes the higher order corrections, which are
unreliable for large $\rho_H$, more important.

Finally, the non-perturbative parameters determined in a shape-function-based
fit  in the P~scheme are also in very good agreement  with the ones found in
the thrust analysis, cf.~Eq.~(\ref{eq:thrust_shaperes_pscheme}), provided
$\alpha_s$ is fixed by the latter:
\begin{eqnarray}
{\rm P \; (udscb): } &&
\begin{tabular}{ll}
$\alpha_{\overline{\mbox{\tiny MS}}}({M_{\rm Z}}) \equiv 0.1068\; \; {\rm fixed} $&  \\
$\lambda_1\bar{\Lambda} = 0.404\pm 0.007 \; \; {\rm GeV}  $ &
$\lambda_2\bar{\Lambda}^2 = 0.080\pm 0.016 \; \; {\rm GeV}^2 $ \\
$b_1 = 0.37 \pm  0.03$  &$b_2 \equiv 0$ \\
$q = 0.46  \pm 0.07 $&
\end{tabular}
\end{eqnarray}
with $\chi^2/{\rm dof}=0.82$ for 99  points. Although 
the total $\chi^2$ is good, the contribution from
the experiments at ${M_{\rm Z}}$ is still larger than 2 per point.
The result for $\lambda_1$ is rather stable with respect to variations of both the upper and lower cuts. The result for $\lambda_2$ turns out to be
stable with respect to the upper cut, but, as in the decay scheme, not with respect to the lower cut. 
Fixing also the shape function to be the one found in the thrust
analysis, the $\chi^2$ increases to $\chi^2/{\rm dof}=1.44$ for 99
points. In this case the $\chi^2$ is acceptable for all data sets
except OPAL at ${M_{\rm Z}}$ and SLD, which contribute to $\chi^2$ 
$84$ for $10$ points and $13$ for $2$ points, respectively.

To summarize the comparison between the heavy-jet mass and the thrust, we find a good
agreement between the two for the first moment of the shape function (or the shift) 
$\lambda_1$ once $\alpha_s$ is fixed. Although $\lambda_2$ is harder to determine,
there is no indication for discrepancy between the observables even at this level. The
agreement is restricted, of course,  to the massless schemes (P, E and D).     In
general, the heavy-jet mass
distribution in the peak region ($\rho_H\lsim 0.1$) is well described by the 
theoretical prediction that uses $\alpha_s$ and the non-perturbative shape
function from the thrust analysis as input.  This supports our theoretical
assumptions,  in particular: using DGE as a basis for power correction
analysis  in the peak region (NLL resummation is insufficient),  the assumption
that the correlation between the hemisphere masses in the peak region is small
(the simple relation between the  thrust and the heavy-jet mass is based on no
correlation)  and, finally, the assumption that the non-perturbative
corrections implied by the renormalon ambiguities dominate. In
addition, we demonstrated how a careful treatment of hadron mass effects can
open the way to quantitative comparison of hadronization effects which are 
parametrized as power corrections.


\section{Conclusions} 

Dressed gluon exponentiation is especially designed to resum large
perturbative  corrections
in differential cross sections in the threshold region, where
power-suppressed corrections are non-negligible. DGE provides the basis for
 parametrization of these power corrections.  
Contrary to the standard Sudakov  resummation techniques, which aim  
at a fixed logarithmic accuracy (such as NLL accuracy), DGE attempts to
reach power accuracy. This does not mean, of course, that the perturbative
calculation itself is correct to power accuracy (it is ambiguous at this level)
but rather that the larger, perturbative corrections are resummed.  These
corrections appear as subleading Sudakov logs that are enhanced by numerical
coefficients, which increase factorially at large orders.

Physically, the dominant perturbative corrections and the power corrections
have the same source: soft gluon emission at large angles. Although both arise
from integrals over the running coupling, one must separate between them, resum
the former and parametrize the latter. A systematic separation is achieved, for
example, by Borel summation.  The sum of the two contributions is physically
meaningful, but each of them separately is ambiguous. Moreover, subleading
Sudakov logs (which are above the minimal term in the expansion) are
parametrically larger than the ambiguous power correction, and their functional
dependence on $Q$ and $\rho$ is distinct.  Having realized this, it is clear
that a meaningful comparison with data  at power accuracy cannot be achieved
with a fixed logarithmic accuracy. DGE is necessary.      

Starting with the DGE calculation of the single-jet mass distribution in the
approximation of independent emission, we expressed both the thrust and the
heavy-jet mass distributions. Although the single-jet mass distribution is
sensitive to non-inclusive  contributions~\cite{Dasgupta:2001sh}, both the
thrust and the heavy-jet mass have a  reduced sensitivity: Abelian
non-inclusive contributions appear first at $\rm N^{(3)}LL$ order while
non-Abelian ones do not appear before NNLL (see section~\ref{sec:applicability}). 

An immediate consequence of the resummation of running-coupling effects by DGE
is that the Sudakov exponent is renormalization-scale invariant within a
given renormalization scheme. This is, of course, in contrast to NLL resummation. Fixing the renormalization scheme requires, in general, to fix
$\Lambda$ as well as the renormalization-group equation for the
coupling. Here $\Lambda$ was {\em unambiguously} fixed as $\bar{\Lambda}$
in the gluon bremsstrahlung scheme~\cite{Catani:1991rr},
accounting for the singular terms in the splitting function at
next-to-leading order, which contribute to the thrust and the heavy-jet mass
at the NLL level.
On the other hand, the renormalization-group equation was {\em  arbitrarily}
chosen as the `t Hooft scheme, in which only the first two coefficients are non-vanishing. Consequently, renormalization-scheme
dependence appears (through terms that are subleading in $\beta_0$) at NNLL
order and beyond.
Let us also recall that since renormalons are resummed
only at the exponent, renormalization-scale dependence eventually appears
when matching with the NLO result.
Numerically these scheme and scale dependence effects are small and they
are sub-dominant\footnote{Estimates have been given in section 5.6 in
\cite{Average_thrust} and in section 5.1 in~\cite{Thrust_distribution}.}
with respect to other theoretical uncertainties which
are discussed below.

As discussed in~\cite{Thrust_distribution}, the DGE formula is
suggestive of specific power corrections. First of all, the appearance of
renormalon ambiguities in the exponent implies that power corrections also
exponentiate, i.e. factorize in the Laplace space. This is compatible with the
shape-function approach of \cite{Korchemsky:2000kp}. To a first approximation,
the non-perturbative correction amounts to a shift of the perturbative
distribution~\cite{Shape_function2,Dokshitzer:1997ew}. Secondly, assuming that
renormalon-related power corrections dominate, intimate information on the
shape function can be deduced from the singularity pattern of (\ref{Borel_nu}).
The most striking property is the  absence of a $\Lambda^2/(\rho Q)^2$ ambiguity
in the exponent. This suggests that, if indeed power corrections follow the
renormalon pattern, the second central  moment of the shape function,
$\lambda_2$, will be suppressed.   

In addition to the non-perturbative corrections that can be inferred from the
renormalon ambiguities, there are also finite hadron mass
effects~\cite{Salam:2001bd} that depend on the way masses  are treated in the
measurement of the variable.  We saw that differences between different
hadron mass schemes are of the same magnitude as the non-perturbative
correction itself. Therefore, it is imperative that one uses the same HMS when
comparing two variables and that this HMS is either insensitive to the masses
or is defined such that the mass corrections for the two observables are
similar. This qualifies the decay scheme as well as the E and the P scheme,
but not the standardly used massive scheme.  We found that, contrary to the
case of average event shapes analysed in~\cite{Salam:2001bd}, the
distributions analysed in different HMS yield very similar  values for
$\alpha_s$. The differences between different HMS can thus be naturally recast
as a redefinition of the non-perturbative parameters.

Analysing the available thrust data, we obtained a best fit value of
$\alpha_{\overline{\mbox{\tiny MS}}}({M_{\rm Z}}) = 0.109\pm 0.001 $. This result
is obtained both when fitting with a shift above the peak region and
when fitting with a shape function. The consistency between the two  procedures
is reassuring. The main uncertainty in this result was found to be the effect
of contributions suppressed by powers of $t$.  It was estimated as $\sim 4$\%
by modifying the logarithm in the resummation from $\ln(1/t)$ to $\ln(1/t-1)$.
The only other important uncertainty is due to the HMS dependence, which was
found to be $\sim 2$\%.  There is, of course, some systematic uncertainty from
the Monte Carlo  model used to transform the data between different schemes,
which we did not investigate. Other uncertainties, which we have investigated,
turn out to be small: $\sim 0.5$\%, i.e. comparable to the propagated
experimental error. This includes: using a shift or a shape function, the form
of the shape function, the fitting range, the
residual renormalization scale dependence~\cite{Thrust_distribution}, and the effect of heavy quarks. The use of the Monte Carlo model in quantifying the latter is not satisfactory. All in
all our estimate for the uncertainty in $\alpha_s$ is  $\sim5$\%.

For the non-perturbative parameters of the shape function, our best fit values
in the decay scheme (with all primary quarks) are $\lambda_1\bar{\Lambda} =
0.605 \pm 0.013$ GeV and $\lambda_2\bar{\Lambda}^2 = 0.002 \pm 0.024$ GeV$^2$.
The same value for $\lambda_1$ is obtained when fitting with shift.  In other
hadron mass schemes the value of $\lambda_1$ can be a factor $\sim 2$ lower. It
is important to remember that the non-perturbative parameters are strongly
correlated to $\alpha_s$. In a given HMS, with a fixed $\alpha_s$, the total
uncertainty in $\lambda_1$ is  less than $\sim 10$\%. We also found a weak
(logarithmic) energy dependence when fitting $\lambda_1$  separately at each
energy. The second central moment of the shape function,
$\lambda_2$, is correlated with both $\alpha_s$ and $\lambda_1$. Nevertheless,
it is worth noting that in the best fit in the decay scheme $\lambda_2$
vanishes, in agreement with the large $\beta_0$ renormalon pattern. This
finding can regarded as a success of the renormalon dominance assumption.
However, one should be aware that it is not shared by other HMS: 
in the P scheme $\lambda_2$ is significantly different from zero. 
Hadron mass effects do not have a ``perturbative'' origin, and therefore they 
certainly do not admit the renormalon power correction pattern.

In the approximation of independent emission, soft gluons are radiated from a
quark antiquark dipole. This naive phase space is useful to calculate the
logarithmically enhanced terms, which are associated with this configuration.
However, it misses kinematic constraints, which are present in the exact
 phase space. As discussed in  section~\ref{sec:applicability},
this shows up as non-logarithmic corrections, which become important away from
the two-jet limit.  By examining our approximation versus the exact NLO
calculation, we found that the thrust distribution is quite insensitive to
such constraints, whereas the  heavy-jet mass~is. The difference between the
heavy-jet mass and the thrust is well understood. The kinematic constraints
imply correlations between the hemisphere masses: specifying the heavy-jet
mass, as opposed to the thrust, puts a stringent constraint on the light jet
mass. As~a~consequence, the logarithmic approximation to the heavy-jet mass
distribution breaks down rather early,  at $\rho_H= 1/6$, and it is not as
accurate as the equivalent calculation of the thrust distribution. Since the
range of $\rho_H$ values where the  resummation formula is valid is rather
narrow, the determination of $\alpha_s$ from this distribution is unreliable:
independent fits yield values of $\alpha_s$ significantly smaller than
the thrust and values of $\lambda_1$ correspondingly larger. We showed that the
distortion of the distribution mainly affects larger values of $\rho_H$. This
is reflected for example in the fact that the best fit value of $\alpha_s$
decreases as the fitting range shifts to the right. Therefore, the comparison
of non-perturbative corrections in the two-jet region  between the heavy-jet
mass and the thrust is still possible.

Since perturbatively the correlation between the hemispheres is significant only
away from the two-jet limit, it is natural to assume that the correlation  in
the peak region is small also on the non-perturbative level.  Under this
assumption, a single non-perturbative shape function, which is associated with
the single-jet mass, incorporates the dominant non-perturbative corrections to
both the thrust and the heavy-jet mass distributions. 

The deviation from this simple scenario was the subject of an  interesting
theoretical study and phenomenological
analysis~\cite{Korchemsky:2000kp,Belitsky:2001ij}.  It was claimed that
correlation between the hemispheres due to non-perturbative soft-gluon emission
is essential for a consistent description of the heavy-jet mass distribution in
the peak region\footnote{Two major differences between the analysis
in~\cite{Korchemsky:2000kp} and ours should be noted: the resummation formula 
applied in~\cite{Korchemsky:2000kp} is the standard NLL one (with an infrared cutoff), and the data for the heavy-jet mass are in the massive scheme.}. 
Ref.~\cite{Belitsky:2001ij} finds that, owing to radiation from one hemisphere into the other (``non-inclusive'' correction) there is a {\em positive} correlation between the jet masses. It should be emphasized that this dynamical
correlation, which is introduced in~\cite{Korchemsky:2000kp} through the shape
function,  has a similar flattening impact on the thrust distribution as on the 
$\rho_H$ distribution. Whereas radiation from
one hemisphere into the other induces a positive correlation between the
hemisphere masses, purely kinematic considerations indicate a {\em negative}
correlation, which is missed by the logarithmic approximation for the heavy-jet mass distribution. As discussed in
section~\ref{sec:applicability}, the latter seems to be 
the dominant feature limiting the accuracy of the calculation.

We assume here that the
correlation between the hemispheres in the peak region is small and, we confront
this assumption with the data. This is done is by using the best fits
of the thrust distribution to determine~$\alpha_s$ and the parameters of the
shape function, and then confront the {\em calculated} heavy-jet mass
distribution with the data. The consistency of the results shows that the
correlation in the peak region is indeed small. An alternative approach we
applied was to perform fits to the heavy-jet mass data, fixing $\alpha_s$ by
the thrust analysis. We found a good agreement between the values of the shift
$\lambda_1$ in the decay and E schemes, and a reasonable agreement in the P
scheme (of course, the values differ by much in the massive scheme).
Repeating the same exercise with the NLL resummation instead of DGE, we found a
strong discrepancy between the thrust and heavy-jet mass results.  The
agreement of $\lambda_1$ between the thrust and the heavy-jet mass  persists
when fitting with a shape function. For the second central moment
$\lambda_2$ the results are somewhat less conclusive, since it is sensitive to
the fitting range and to the functional form of the shape function.
Nevertheless, there is no indication for discrepancy between the observables
even at this level. 

It is of great interest to push the accuracy of the calculations further, e.g.
by matching with a full NNLO result when it becomes avaliable. Some interesting
questions could then be addressed in more detail. This includes, in particular,
the consistency of the higher moments of the shape function with the renormalon
pattern~\cite{Thrust_distribution}, the correlation between the
hemispheres~\cite{Korchemsky:2000kp,Belitsky:2001ij}, and hadron mass
effects~\cite{Salam:2001bd}.

\section*{Acknowledgements}

We would like to thank Gregory Korchemsky, Gavin Salam and Daniel Wicke for
helpful discussions.


\begin{thebibliography}{99}

\bibitem{Sterman:1977wj}
G.~Sterman and S.~Weinberg,
{\em Phys. Rev. Lett.}   {\bf 39} (1977) 1436.

\bibitem{Catani:1991kz}
S.~Catani, G.~Turnock, B.~R.~Webber and L.~Trentadue,
{\em Phys. Lett.}   {\bf B263} (1991) 491.

\bibitem{Catani:1993ua}
S.~Catani, L.~Trentadue, G.~Turnock and B.~R.~Webber,
{\em Nucl. Phys.}  {\bf B407} (1993) 3.

\bibitem{Catani:1992jc}
S.~Catani, G.~Turnock and B.~R.~Webber,
{\em Phys. Lett.}   {\bf B295} (1992) 269.


\bibitem{Catani:1998sf}
S.~Catani and B.~R.~Webber,
{\em Phys. Lett.}  {\bf  B427} (1998) 377
[hep-ph/9801350].


\bibitem{Dokshitzer:1998kz}
Y.~L.~Dokshitzer, A.~Lucenti, G.~Marchesini and G.~P.~Salam,
{\em JHEP} {\bf 9801} (1998) 011
[hep-ph/9801324].


\bibitem{Average_thrust}
E.~Gardi and G.~Grunberg,
{\em JHEP} {\bf 9911} (1999) 016
[hep-ph/9908458].

\bibitem{Gardi:2000yh}
E.~Gardi,
{\em JHEP} {\bf 0004} (2000) 030
[hep-ph/0003179].


\bibitem{Thrust_distribution}
E.~Gardi and J.~Rathsman,
{\em Nucl. Phys.}  {\bf B609} (2001) 123
[hep-ph/0103217].



\bibitem{Contopanagos:1994yq}
H.~Contopanagos and G.~Sterman,
{\em Nucl. Phys.}  {\bf B419} (1994) 77
[hep-ph/9310313].

\bibitem{Manohar:1995kq}
A.~V.~Manohar and M.~B.~Wise,
{\em Phys. Lett.}   {\bf B344} (1995) 407
[hep-ph/9406392].

\bibitem{Webber:1994cp}
B.~R.~Webber,
{\em Phys. Lett.}   {\bf B339} (1994) 148
[hep-ph/9408222].

\bibitem{Korchemsky:1995is}
G.~P.~Korchemsky and G.~Sterman,
{\em Nucl. Phys.}  {\bf B437} (1995) 415
[hep-ph/9411211].


\bibitem{Shape_function2} G.P. Korchemsky and G. Sterman,
Proc. 30th Rencontres de Moriond, {\em
QCD and high energy hadronic interactions}, Les Arcs, Savoie,
France, 1995, ed. J. Tran Thanh Van (Editions
Fronti\`eres, Gif-sur-Yvette, 1995), p. 383 [hep-ph/9505391].


\bibitem{Shape_function3}
G.~P.~Korchemsky, {\em Shape functions and power corrections to
the event shapes}, hep-ph/9806537.

\bibitem{Korchemsky:1999kt}
G.~P.~Korchemsky and G.~Sterman,
{\em Nucl. Phys.}  {\bf B555} (1999) 335
[hep-ph/9902341].

\bibitem{Korchemsky:2000kp}
G.~P.~Korchemsky and S.~Tafat,
{\em JHEP} {\bf 0010} (2000) 010
[hep-ph/0007005].

\bibitem{Belitsky:2001ij}
A.~V.~Belitsky, G.~P.~Korchemsky and G.~Sterman,
{\em Phys. Lett.}   {\bf B515} (2001) 297
[hep-ph/0106308].

\bibitem{Dokshitzer:1995zt}
Y.~L.~Dokshitzer and B.~R.~Webber,
{\em Phys. Lett.}   {\bf B352} (1995) 451
[hep-ph/9504219].

\bibitem{Akhoury:1995sp}
R.~Akhoury and V.~I.~Zakharov,
{\em Phys. Lett.}   {\bf B357} (1995) 646
[hep-ph/9504248].


\bibitem{Dokshitzer:1998pt}
Y.~L.~Dokshitzer, A.~Lucenti, G.~Marchesini and G.~P.~Salam,
{\em JHEP} {\bf 9805} (1998) 003
[hep-ph/9802381].

\bibitem{DMW}
Y.~L.~Dokshitzer, G.~Marchesini and B.~R.~Webber,
{\em Nucl. Phys.}  {\bf B469} (1996) 93
[hep-ph/9512336].

\bibitem{Dokshitzer:1997ew}
Y.~L.~Dokshitzer and B.~R.~Webber,
{\em Phys. Lett.}   {\bf B404} (1997) 321
[hep-ph/9704298].

\bibitem{Salam:2001bd}
G.~P.~Salam and D.~Wicke,
{\em JHEP} {\bf 0105} (2001) 061
[hep-ph/0102343].

\bibitem{DGE}
E.~Gardi,
{\em Nucl. Phys.} {\bf B622} (2002) 365 
[hep-ph/0108222].

\bibitem{BLM}
S.~J.~Brodsky, G.~P.~Lepage and P.~B.~Mackenzie,
{\em Phys. Rev.}   {\bf D28} (1983) 228. \\
G.~P.~Lepage and P.~B.~Mackenzie,
{\em Phys. Rev.}   {\bf D48} (1993) 2250
[hep-lat/9209022].


\bibitem{Beneke:1995qe}
M.~Beneke and V.~M.~Braun,
{\em Phys. Lett.}   {\bf B348} (1995) 513
[hep-ph/9411229].

\bibitem{Ball:1995ni}
P.~Ball, M.~Beneke and V.~M.~Braun,
{\em Nucl. Phys.}  {\bf B452} (1995) 563
[hep-ph/9502300].

\bibitem{BGGR}
S.~J.~Brodsky, E.~Gardi, G.~Grunberg and J.~Rathsman,
{\em Phys. Rev.}   {\bf D63} (2001) 094017
[hep-ph/0002065].

\bibitem{Contopanagos:1997nh}
H.~Contopanagos, E.~Laenen and G.~Sterman,
{\em Nucl. Phys.}  {\bf B484} (1997) 303
[hep-ph/9604313].

\bibitem{MovillaFernandez:2001ed}
P.~A.~Movilla Fernandez, S.~Bethke, O.~Biebel and S.~Kluth,
{\em Eur. Phys. J. C} {\bf 22} (2001) 1 [hep-ex/0105059].

\bibitem{PDG}
D.~E.~Groom {\it et al.}  [Particle Data Group Collaboration],
{\em Eur. Phys. J.}  {\bf C15} (2000) 1, and 2001 off-year partial update for the 2002 edition available on the PDG
WWW pages (URL: http://pdg.lbl.gov/).

\bibitem{Catani:1991rr}
S.~Catani, B.~R.~Webber and G.~Marchesini,
{\em Nucl. Phys.}  {\bf B349} (1991) 635.



\bibitem{Brown:1992ic}
L.~S.~Brown and L.~G.~Yaffe,
{\em Phys. Rev.}   {\bf D45} (1992) 398.

\bibitem{Brown:1992pk}
L.~S.~Brown, L.~G.~Yaffe and C.~Zhai,
{\em Phys. Rev.}   {\bf D46} (1992) 4712
[hep-ph/9205213].

\bibitem{Beneke:1993ch}
M.~Beneke,
{\em Nucl. Phys.}  {\bf B405} (1993) 424.

\bibitem{Grunberg:1993hf}
G.~Grunberg,
{\em Phys. Lett.}   {\bf B304} (1993) 183.

\bibitem{Nason:1995hd}
P.~Nason and M.~H.~Seymour,
{\em Nucl. Phys.}  {\bf B454} (1995) 291
[hep-ph/9506317].

\bibitem{Dasgupta:2001sh}
M.~Dasgupta and G.~P.~Salam,
{\em Phys. Lett.}   {\bf B512} (2001) 323
[hep-ph/0104277].

\bibitem{EVENT2} S.~Catani and M.H.~Seymour,
{\em Phys. Lett.}  {\bf B378} (1996) 287;
{\em Nucl. Phys.} {\bf B485} (1997) 291;
{\tt http://hepwww.rl.ac.uk/theory/seymour/nlo/ }

\bibitem{Clavelli:1981yh}
L.~Clavelli and D.~Wyler,
{\em Phys. Lett.}   {\bf B103} (1981) 383.

\bibitem{PYTHIA}
T.~Sj\"ostrand, P.~Eden, C.~Friberg, L.~L\"onnblad, G.~Miu, S.~Mrenna and E.~Norrbin,
{\em Comput. Phys. Commun.}  {\bf 135} (2001) 238
[hep-ph/0010017].

\bibitem{Lonnblad:1992tz}
L.~Lonnblad,
{\em Comput. Phys. Commun.}  {\bf 71} (1992) 15.

\bibitem{HERWIG}
G.~Corcella {\it et al.},
{\em JHEP} {\bf 0101} (2001) 010
[hep-ph/0011363].

\bibitem{TASSO}
W.~Braunschweig {\it et al.}  [TASSO Collaboration],
{\em Z. Phys.}    {\bf  C47} (1990) 187.

\bibitem{TASSOMASS}
W.~Braunschweig {\it et al.}  [TASSO Collaboration],
{\em Z. Phys.}    {\bf C45} (1989) 11.
\bibitem{JADE}
P.~A.~Movilla Fernandez, O.~Biebel, S.~Bethke, S.~Kluth and P.~Pfeifenschneider
                  [JADE Collaboration],
{\em Eur. Phys. J.}  {\bf C1} (1998) 461
[hep-ex/9708034].

\bibitem{ALEPH91}
R.~Barate {\it et al.}  [ALEPH Collaboration],
{\em Phys. Rep.} {\bf 294} (1998) 1.

\bibitem{DELPHI91}
P.~Abreu {\it et al.}  [DELPHI Collaboration],
{\em Eur. Phys. J.}  {\bf C14} (2000) 557
[hep-ex/0002026].

\bibitem{L391}
B.~Adeva {\it et al.}  [L3 Collaboration],
{\em Z. Phys.}    {\bf C55} (1992) 39.

\bibitem{OPAL91}
P.~D.~Acton {\it et al.}  [OPAL Collaboration],
{\em Z. Phys.}    {\bf C59} (1993) 1;

\bibitem{SLD91}
K.~Abe {\it et al.}  [SLD Collaboration],
{\em Phys. Rev.}   {\bf D51} (1995) 962
[hep-ex/9501003].

\bibitem{ALEPH133}
D.~Buskulic {\it et al.}  [ALEPH Collaboration],
{\em Z. Phys.}    {\bf C73} (1997) 409.

\bibitem{DELPHI133_161_172_183}
P.~Abreu {\it et al.}  [DELPHI Collaboration],
{\em Phys. Lett.}   {\bf B456} (1999) 322.


\bibitem{OPAL133}
G.~Alexander {\it et al.}  [OPAL Collaboration],
{\em Z. Phys.}    {\bf C72} (1996) 191.


\bibitem{OPAL161}
K.~Ackerstaff {\it et al.}  [OPAL Collaboration],
{\em Z. Phys.}   {\bf  C75} (1997) 193.


\bibitem{OPAL172_183_189}
G.~Abbiendi {\it et al.}  [OPAL Collaboration],
{\em Eur. Phys. J.}  {\bf C16}, 185 (2000)
[hep-ex/0002012].


\end{thebibliography}
\end{document}